\documentclass[aps,superscriptaddress]{revtex4}
\usepackage{amssymb,amsmath,epsfig}
\usepackage{subfigure}
\usepackage{color}
\usepackage[colorlinks=true, pdfstartview=FitV, linkcolor=blue, citecolor=red, urlcolor=magenta, breaklinks=true]{hyperref}
\usepackage{graphicx}
\usepackage{latexsym}
\usepackage{amsfonts}
\usepackage{multirow}
\usepackage{verbatim}
\graphicspath{{Graf/}}

\begin{document}

\title{Absorption, scattering, quasinormal modes and shadow by canonical acoustic black holes in Lorentz-violating background}

\author{J. A. V. Campos}\email{joseandrecampos@gmail.com}
\affiliation{Departamento de F\'{\i}sica, Universidade Federal de Campina Grande
Caixa Postal 10071, 58429-900 Campina Grande, Para\'{\i}ba, Brazil}

\author{M. A. Anacleto}\email{anacleto@df.ufcg.edu.br}
\affiliation{Departamento de F\'{\i}sica, Universidade Federal de Campina Grande
Caixa Postal 10071, 58429-900 Campina Grande, Para\'{\i}ba, Brazil}

\author{F. A. Brito}\email{fabrito@df.ufcg.edu.br}
\affiliation{Departamento de F\'{\i}sica, Universidade Federal de Campina Grande
Caixa Postal 10071, 58429-900 Campina Grande, Para\'{\i}ba, Brazil}
\affiliation{Departamento de F\'isica, Universidade Federal da Para\'iba, 
Caixa Postal 5008, 58051-970 Jo\~ao Pessoa, Para\'iba, Brazil}

\author{E. Passos}\email{passos@df.ufcg.edu.br}
\affiliation{Departamento de F\'{\i}sica, Universidade Federal de Campina Grande
Caixa Postal 10071, 58429-900 Campina Grande, Para\'{\i}ba, Brazil}


\begin{abstract}
In the present work, we study the scattering for a black hole described by the canonical acoustic metric with Lorentz violation using asymptotic and numerical methods. In this scenario, we also check the effects of quasinormal modes and the acoustic shadow radius. In the
eikonal limit the relationship between the shadow radius and the real part of the quasinormal frequency is preserved.
\end{abstract}

 \maketitle
\section{Introduction}
One of the most fascinating predictions of general relativity is black holes being solutions to the classic Einstein field equations in a vacuum. Today in astrophysics, the study of black holes is growing motivated by the experimental results obtained such as the detection of gravitational waves by the LIGO-VIRGO collaboration \cite{scientific2016tests,abbott2017gw170817}, the first image of a black hole located in the center of the galaxy M87 and more recently the  incredible image of the black hole at the center of our galaxy, Sagittarius A* by the \textit{Event horizon Telescope} (EHT)\cite{akiyama2019firstResultsIV,EventHorizonTelescope:2019ggy, akiyama2022first}. 
Another way of experimentally studying the physics of black holes would be to reproduce analogous models in the laboratory. Unruh \cite{unruh1981experimental} was one of the pioneers to suggest an acoustic black hole model 
to investigate the physical properties of black holes, such as Hawking radiation. 
Later, several analogous models of gravity were developed to reproduce the effects of black holes.
{It is well known that an acoustic metric can be constructed considering a moving fluid reaching a speed greater than the local speed of sound, and thus generating an acoustic horizon for the formation of an acoustic black hole.}
{In addition, relativistic acoustic black holes have been generated from the Abelian Higgs model~\cite{ge2010acoustic,ge2019acoustic} as well as in the Lorentz-violating~\cite{anacleto2010acoustic,anacleto2011superresonance,Anacleto:2023ali} and noncommutative~\cite{Anacleto:2011bv,Anacleto:2012du} background (see also~\cite{Anacleto:2021nhm,Anacleto:2013esa}). 
Furthermore, the metric of a background acoustic black hole violating Lorentz has been found by considering terms that violate Lorentz symmetry in the total Lagrangian of the Abelian Higgs model~\cite{anacleto2010acoustic}.}
{Thus, by deriving acoustic black hole metrics from the Lorentz symmetry-breaking Abelian Higgs model, this configuration affects fluid fluctuations and leads to several consequences, for instance the Hawking temperature is directly affected by the Lorentz violation term~\cite{anacleto2010acoustic}.}
For Kerr black hole analogues the Lorentz violation term affects the rate of mass loss and increases the  superresonance phenomenon \cite{anacleto2011superresonance}. 
{The motivation for deriving an acoustic metric from the Abelian Higgs model (that describes high energy physics) with Lorentz symmetry breaking term in the Lagrangian is the fact that in high energy physics, the strong violation of Lorentz symmetry as well as the quark gluon plasma (QGP) can occur together.}
{Therefore, it is of interest to look for acoustic black holes in a QGP fluid with Lorentz symmetry breaking in this regime. In addition, acoustic phenomena in QGP matter can be found in~\cite{Casalderrey-Solana:2004fdk, Das:2020zah} and acoustic black holes in a plasma fluid can be consulted in~\cite{GarciadeAndrade:2008kz,Ditta:2023lny}.}

Analogous models can also simulate effects such as quasinormal modes coming from black holes. The study of disturbances in black holes began with the work of Regge and Wheeler \cite{regge1957stability} analyzing the instability of the Schwarzschild black hole. These disturbances in the vicinity of black holes evolve with peculiar characteristics called quasinormal modes. The frequencies and the damping time of these signals depend only on the black hole parameters like mass, charge and angular momentum \cite{kokkotas1999quasi}. The quasinormal modes in acoustic black holes can be found for both (2+1) and (3+1) dimensions \cite{visser1998acoustic, berti2004quasinormal, cardoso2004quasinormal}. In \cite{cardoso2004quasinormal} the issue of dimensional instability of a rotating draining bathtub acoustic black hole is verified. Although the WKB method shows a change in the sign of the imaginary part indicating instability, it was numerically verified that there is no instability for small disturbances.

Currently, quasinormal modes can be studied experimentally by detecting gravitational waves arising from the merger of black holes and neutral stars such as those presented by the LIGO-VIRGO collaboration. Obtaining the quasinormal modes theoretically makes use of approximate analytical or numerical methods to solve the perturbation equations. The most common one makes use of the WKB approximation with higher-order corrections \cite{schutz1985black, iyer1987black, konoplya2003quasinormal}. In this article, we will use the sixth-order WKB approximation implemented by Konoplya. Another way to obtain the quasinormal frequencies is using numerical methods. One of the methods that present very satisfactory results was developed by Leaver \cite{leaver1985analytic, leaver1990quasinormal}, but in our case this method became complicated due to the high number of terms that connect the relation of recurrence.
In the present work, we will verify the effects of Lorentz symmetry breaking on an acoustic black hole metric in (3+1) dimensions, by studying the behavior of the absorption and scattering cross sections as well as the quasinormal modes, verifying the validity of the relation between the shadow of the black hole and the quasinormal frequencies in the eikonal limit.

The paper is organized as follows. In Sec. \ref{S2} we present the metric for an acoustic black hole in (3+1) dimensions with a term derived from the breaking of the Lorentz symmetry. In Sec. \ref{S3} we analyze the differential scattering and absorption cross sections. We studied the effects of the Lorentz violation term at low and high frequencies using the geodesic method and partial wave analysis for the acoustic metric. We extend the scattering study by verifying the results numerically. In Sec. \ref{S4} we introduce the study of quasinormal modes, verifying the behavior of the real and imaginary part of the quasinormal frequency as well as the evolution in the temporal domain. We also discuss the shadow of a canonical acoustic black hole with a correction term due to Lorentz's violation. Finally, in Sec. \ref{S6} we make our conclusions.

\section{The canonical acoustic black hole}
\label{S2}
{ In this section, we succinctly show in a few steps the derivation of the metric of a canonical acoustic black hole 
in the Lorentz-violating background obtained in~\cite{anacleto2010acoustic}. 
For this purpose, we consider the Abelian Higgs model with Lorentz symmetry breaking implemented in the scalar sector of the theory, which has been constructed in~\cite{anacleto2010acoustic}.}
Hence, the Lagrangian for the Lorentz-violated Abelian Higgs model is given by~\cite{anacleto2010acoustic,anacleto2011superresonance}
\begin{eqnarray}
\mathcal{L} = -\dfrac{1}{4}F_{\mu\nu}F^{\mu\nu} + |D_{\mu}\phi|^{2} + m^{2}|\phi|^{2} - b|\phi|^{4} + k^{\mu\nu}D_{\mu}\phi^{*}D_{\nu}\phi,
\end{eqnarray}
{where $F_{\mu\nu} = \partial_{\mu}A_{\nu} -\partial_{\nu}A_{\mu}$ is the field strength tensor, $A_{\mu}$ is the gauge field, $D_{\mu}\phi = \partial_{\mu}\phi - ieA_{\mu}\phi$ is the covariant derivative, $\phi$ is the complex scalar field  }
and $k^{\mu\nu}$ is a constant symmetric tensor implementing Lorentz symmetry breaking. 
{Here, we highlight that the upper limit for the components of $k_{\mu\nu}$ are $k_{00} \leq 3.6 \times 10^{-8}$ \cite{mohr2008codata} and
$tr(k_{ij} ) \leq 3 \times 10^{-6}$ in relativistic and non-relativistic BEC theory~\cite{Casana:2011bv}.  
For the sake of simplicity}, we reduce the ten components of the tensor $k_{\mu \nu}$ into two independent components by choosing the following entries $k_{ii} = k_{00}\equiv \beta$ and $k_{0i}=k_{ij}\equiv \alpha$ {(although for our calculations this produces a scenario rich enough, other effects could also be obtained by making other choices)} such that the tensor is given by the form
\begin{eqnarray}
k_{\mu\nu} = \left[{\begin{array}{cccc}\beta & \alpha & \alpha & \alpha\\
\alpha & \beta & \alpha & \alpha\\
\alpha & \alpha & \beta & \alpha\\
\alpha & \alpha & \alpha & \beta\\
\end{array}}\right] \qquad (\mu, \nu = 0, 1, 2, 3),
\label{tensor}
\end{eqnarray}
being $\alpha$ and $\beta$ real parameters with a magnitude of the order $k_{00}$. 
We assume all components with magnitude around the shortest limit $k_{00} \leq 3.6 \times 10^{-8}$. 
{The choice of tensor $k_{\mu\nu}$ in (\ref{tensor}) has the effect of modifying boosts when we consider the situation $k_{00}=k_{ii}=\beta\neq 0$ and $k_{0i}=k_{ij}=\alpha=0 $ and the birefringence phenomenon arises by choosing $k_{00}=k_{ii}\equiv\beta=0$ and $k_{0i}=k_{ij}\equiv\alpha\neq 0$ (see~\cite{anacleto2010acoustic,anacleto2011superresonance} for more details).}
Next, we will analyze the (3+1)-dimensional acoustic metric describing a canonical acoustic black hole for the specific case, where
$\beta = 0$ and $\alpha \neq 0$. In this way, we intend to describe the effects due to the presence of the Lorentz symmetry breaking for absorption and differential scattering cross sections and the quasinormal frequencies.

In previous studies \cite{anacleto2010acoustic} were found acoustic metrics in two and three spatial dimensions, describing acoustic black holes. Other studies \cite{anacleto2011superresonance, anacleto2019quantum} focused on the effects of superresonance and quantum corrections on flat acoustic metrics.
In the present work, we focus on the (3 + 1)-dimensional case to obtain a metric for a canonical acoustic black hole in the Lorentz-violating background. 

{Now to obtain the metric of a canonical acoustic black hole in the Lorentz violating background, we adopt the following steps: 
In the first step, we decompose the scalar field as $\phi = \sqrt{\rho(x,t)}\exp{iS(x,t)}$ in the original Lagrangian, such that
\begin{eqnarray}
    \mathcal{L}& =& -\dfrac{1}{4}F_{\mu\nu}F^{\mu\nu} + \rho\partial_{\mu}S\partial^{\mu}S + 2e\rho A_{\mu}\partial^{\mu}S + e^{2}\rho A_{\mu}A^{\mu}+m^{2}\rho - b\rho^{2} 
    \nonumber\\
   & + & k^{\mu\nu}\rho\left(\partial_{\mu} S\partial_{\nu}S - 2eA_{\mu}\partial_{\nu}S + eA_{\mu}A_{\nu}\right)+\dfrac{\rho}{\sqrt{\rho}}\left(\partial_{\mu}\partial^{\mu}+ k^{\mu\nu}\partial_{\mu}\partial_{\nu}\right)\sqrt{\rho}.
\end{eqnarray}
In the second step, we linearize the equations of motion around the background $(\rho_{0},S_{0})$, with $\rho = \rho_{0}+\rho_{1}$, 
$S = S_{0} + \Phi$ and keeping the vector field $A_{\mu}$ unchanged we find the equation of motion for a linear acoustic perturbation $\Phi$ given by a Klein-Gordon equation in curved space $\dfrac{1}{\sqrt{-g}}\partial_{\mu}\left(\sqrt{-g}g^{\mu\nu}\partial_{\nu}\right)\Phi = 0$, 
where $g_{\mu\nu}$  just represent the metric of an acoustic black hole in the Lorentz violating background given explicitly in the equation below.}
Then, by using the aforementioned conditions for the tensor \eqref{tensor}, i.e., $\beta = 0$ and $\alpha \neq 0$, the line element of the acoustic black hole in the Lorentz-violating background in the non-relativistic limit $(v^{2 }<<c_{s}^{2})$ 
{is given by the following acoustic metric~\cite{anacleto2010acoustic}}
\begin{eqnarray}
ds^{2}= \dfrac{b \rho_{0}}{2c_{s}}\left\lbrace-\dfrac{\left[(1 + \alpha)c_{s}^{2} - v_{r}^{2}\right]}{\sqrt{1+\alpha}} dt^{2}  + \dfrac{(1+\alpha)c_{s}^{2}}{\sqrt{1+\alpha}\left[(1+\alpha)c_{s}^{2} - v_{r}^{2}\right]}dr^{2} + \dfrac{\sqrt{1+\alpha c_{s}^{2}}(1 - \alpha v_{r})}{\sqrt{1+\alpha}}r^{2}\left( d\theta^{2} + \sin^{2}\theta d\varphi^{2}\right)\right\rbrace.
\label{dscano}
\end{eqnarray}
{In the sequence to obtain the metric of a canonical acoustic black hole, we consider an incompressible fluid with spherical symmetry. Hence, the density $\rho$ is position-independent, the continuity equation implies that $v \sim 1/r^{2}$ for the fluid velocity field and the sound speed $c_s$ is also a
constant.}
We can define the radial fluid speed as in the form 
\begin{eqnarray}
 v_{r} = c_{s}\frac{r_{0}^{2}}{r^{2}},   
\end{eqnarray} 
where $r_0$ is the radius of the event horizon of the canonical acoustic black hole.
Now, assuming $c_{s} = 1$ and for convenience doing the following redefinition $\alpha \to  2\alpha$ we have the following line element
\begin{eqnarray}
ds^{2}=-(1 + \alpha)\left(1-\frac{r_{h}^{4}}{r^{4}}\right)dt^{2} + (1 + \alpha)^{-1}\left(1-\frac{r_{h}^{4}}{r^{4}}\right)^{-1}dr^{2} + \rho(r)^{2}\left( d\theta^{2} + \sin^{2}\theta d\varphi^{2}\right),
\label{ds}
\end{eqnarray}
where 
\begin{eqnarray}
 r_{h} = \frac{r_{0}}{\sqrt{\alpha + 1}}, 
\end{eqnarray}
is the corrected horizon radius and  
\begin{eqnarray}
\rho(r) = \sqrt{r^{2} - 2\alpha r_{h}^{2}},
\end{eqnarray}
being the appropriate radial coordinate. 
Hence, the coordinate $\rho$ is related to $r_h$ as follows
\begin{eqnarray}
\rho_h = \rho(r_h) = r_h\sqrt{1 - 2\alpha} \approx r_{0}\left(1 - \dfrac{3\alpha}{2} + \dfrac{3\alpha^{2}}{8} + \cdots \right). 
\end{eqnarray}
Which is the suitable horizon radius.

In the following sections we will use the line element above to study the scattering effects and the quasinormal frequencies for the canonical acoustic black hole with Lorentz symmetry breaking.
 
\section{Absorption and scattering cross section}
\label{S3}
In this section, we will explore the scattering and absorption processes by a canonical acoustic black hole in the Lorentz violating background. Initially, we determine the classical scattering through the null geodesic approach. Then, using the partial wave method we compute the differential scattering and absorption cross sections. 

\subsection{Null geodesics and classical scattering}

We can obtain classical scattering by studying geodesic scattering and in the short wavelength limit it is possible to obtain acceptable results for the differential scattering cross section~\cite{collins1973elastic}.
The equations of motion for a particle in an acoustic metric were studied in \cite{dolan2009scattering,oliveira2010absorption}. Here we focus on the study of Lorentz-violated canonical acoustic metric, analyzing the behavior of geodesic lines by numerically solving orbital equations.

\begin{figure}[!htb]
    \centering
    \includegraphics[scale=0.40]{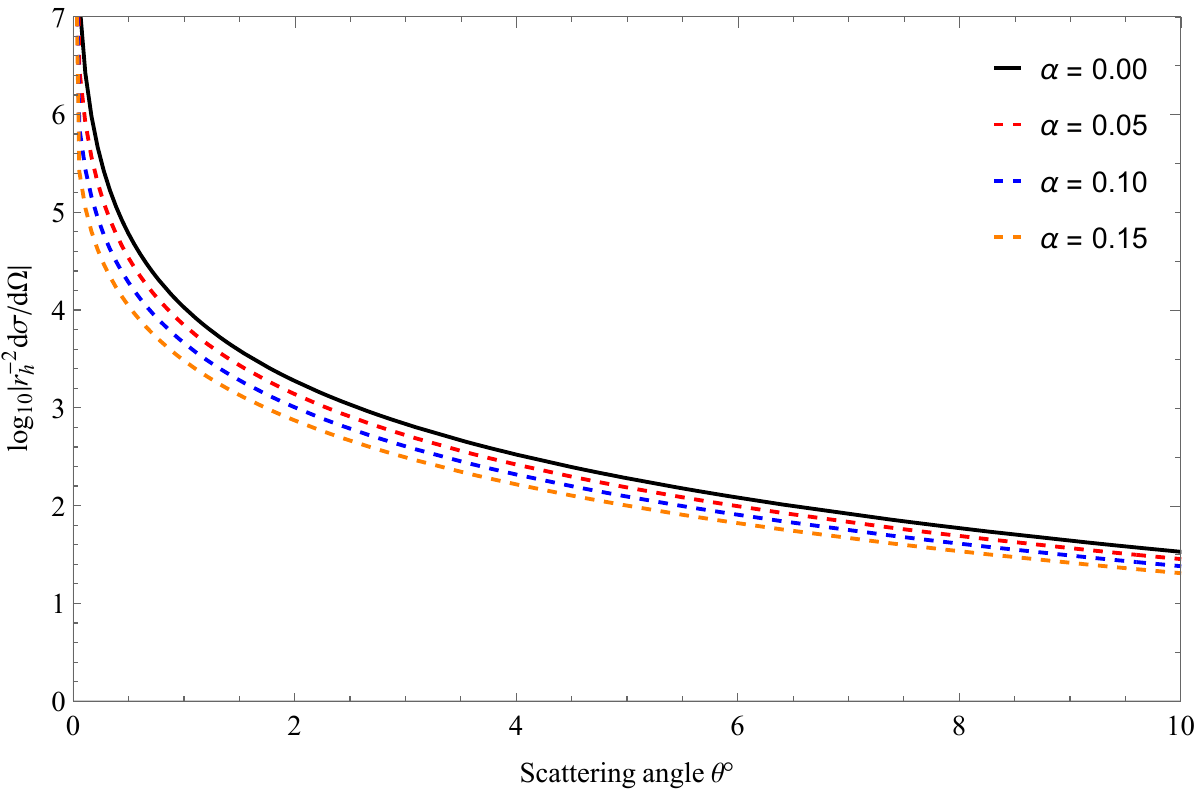} 
    \caption{\footnotesize{{Cross section of classical differential scattering as a function of scattering angle. We can see the influence of the parameter $\alpha$ on the scattering at small angles.}}}
    \label{figScattClas}
\end{figure}

We can find the geodesics from equation \eqref{ds} such that by taking a Lagrangian of the form $\mathcal{L} \equiv \dfrac{1}{2}g_{\mu\nu}\dot{x}^{\mu}\dot{x}^{\nu}$, we have the following 
\begin{equation}
2\mathcal{L} = -(\alpha + 1)\left(1-\frac{r_{h}^{4}}{r^{4}}\right)\dot{t}^{2} + (\alpha + 1)^{-1}\left(1-\frac{r_{h}^{4}}{r^{4}}\right)^{-1}\dot{r}^{2} + \rho^{2}(r)\left(\dot{\theta}^{2} + \sin^{2}\theta\dot{\phi}^{2}\right),
\label{elidot}
\end{equation} 
where the ``." is the derivative concerning the affine parameter.
We are interested in the path of a light ray moving in the equatorial plane $\theta = \pi/2$.
Under these conditions, the metric \eqref{elidot} is independent of $t$ and $\phi$. Thus, two equations are sufficient to describe the movement of a ray of light, we can assemble a system with these equations that give rise to two constants of geodesic motion $E$ and $L$, which correspond to energy and angular momentum respectively, so we get the following constants
\begin{equation}
E = (\alpha + 1)\left(1-\frac{r_{h}^{4}}{r^{4}}\right)\dot{t},\qquad \quad L = \rho^{2}(r)\dot{\phi}.
\label{EL}
\end{equation}
For null geodesics, we have $g_{\mu\nu}\dot{x}^{\mu}\dot{x}^{\nu} = 0$ and using equation \eqref{EL}, we can write the equation of ``energy"
\begin{equation}
\dot{r}^{2}  + \dfrac{(\alpha + 1)\left(r^{4}-r_{h}^{4}\right)L^{2}}{r^{4}\left(r^{2}-2\alpha r_{h}^{2}\right)} = E^{2}.
\label{eqEner}
\end{equation}
Introducing a new variable $u = 1/r$, we can write the orbital equation as follows
\begin{eqnarray}
&\left(\dfrac{du}{d\phi}\right)^{2} = \left(1-2\alpha r_{h}^{2}u^{2}\right)^{2}\left[\dfrac{1}{b^2} - \dfrac{(\alpha+1)\left(1-r_{h}^{4}u^{4}\right)u^{2} }{\left(1-2\alpha r_{h}^{2}u^{2} \right)}\right] \label{eqD1},\\
&\dfrac{1}{(\alpha + 1)}\dfrac{d^{2}u}{d\phi^{2}} + u = 3r_{h}^{4}u^{5}- 4\alpha r_{h}^{6}u^{7} - 4\alpha r_{h}^{2}u\left[ \dfrac{\left(1-2\alpha r_{h}^{2}u^{2}\right)}{(\alpha +1)b^{2}} - \left(1-r_{h}^{4}u^{4}\right)u^{2}\right],
\label{eqD2}
\end{eqnarray}
where $b = L/E$ is the impact parameter. The second-order differential equation is a version of Binet's equation. In the critical case, we have two conditions, the first is $du/d\phi = 0$. Thus, we get the critical impact parameter given by 
\begin{eqnarray}
 b_{c}^{2} = \dfrac{1-2\alpha r_{h}^{2}u_{c}^{2}}{(1+\alpha)\left(1 - r_{h}^{4}u_{c}^{4}\right)u_{c}^{2}},
\label{b_rcritico}
\end{eqnarray}
being $u_{c} = 1/r_{c}$. The classical scattering cross section is simply the area for a circle of radius corresponding to the critical impact parameter $b_{c}$, i.e., $\sigma_{abs}^{hf} = \pi b_{c}^{2}$. From the second condition, $d^{2}u/d\phi^2 = 0$, we obtain the equation for the critical radius
\begin{eqnarray}
\qquad r_{c}^{6}-3r_{h}^{4}r_{c}^{2}+4\alpha r_{h}^{6} = 0.
\label{c_rcritico}
\end{eqnarray}
Note that both the impact parameter and the critical radius will be modified by the correction parameter. 
Equation \eqref{c_rcritico} can be written, with $r^2_c=z$, in the form
\begin{eqnarray}
z^3 - 3r^4_h z + 4\alpha r^6_h=0,
\end{eqnarray}
whose roots are given by~\cite{Anacleto:2020efy, visser2013area}
\begin{eqnarray}
z=2r^2_h\sin\left[\frac{1}{3}\sin^{-1}\left(2\alpha  \right)  + \epsilon\frac{2\pi}{3}\right], \qquad 
\epsilon \in \{0,\pm 1 \}.
\end{eqnarray}
Hence, for $\epsilon=1,0,-1$, up to first order in $\alpha$, we find the following roots:
\begin{eqnarray}
&&r^2_{c,1}=r^2_h \left( \sqrt{3} - \frac{2\alpha}{3} + \cdots \right),
\label{rc1}
\\
&&r^2_{c,0}= \frac{4\alpha r_{h}^{2}}{3} + \cdots, 
\\
&&r^2_{c,-1}=-r^2_h \left(\sqrt{3} + \frac{2\alpha}{3} + \cdots \right).
\end{eqnarray}
Now substituting \eqref{rc1} into \eqref{b_rcritico}, we find
\begin{eqnarray}
 \sigma_{abs}^{hf} &=& \pi b_{c}^{2}
 =\pi\dfrac{r_{c,1}^{4}(r_{c,1}^{2} - 2\alpha r_{h}^{2})}{(1+\alpha)^{2}(r_{c,1}^{4}-r_{h}^{4})},
 \\
 &=& \dfrac{3\sqrt{3}\pi r_{h}^{2}}{2} - \dfrac{3\pi\alpha(2+\sqrt{3})r_{h}^{2}}{2} + \dots + O(\alpha^2).
 \label{sighf}
\end{eqnarray}
So that when $\alpha = 0$, we have 
\begin{eqnarray}
 r_c=(\sqrt{3}r^2_h)^{1/2},   
\end{eqnarray}
and 
\begin{eqnarray}
\sigma_{abs}^{hf} = \pi b_{c}^{2}
=\pi r^2_c \left[1-\frac{r^4_h}{r^4_c}\right]^{-1}=\frac{3\sqrt{3}\pi r^2_h}{2}.   
\end{eqnarray}
Next, solving numerically the equations \eqref{eqD1} and \eqref{eqD2}, we can obtain the behavior of the geodesic lines, verifying the critical impact parameter and the scattering effect. We can also solve the orbital equation \eqref{eqD1} to obtain the geodesic deflection angle, given by \cite{collins1973elastic} 
\begin{eqnarray}
\Theta(b) = 2\int^{u_{1}}  \dfrac{1}{\left(1-2\alpha r_{h}^{2}u^{2}\right)} \left[\dfrac{1}{b^2} - \dfrac{(\alpha +1)\left(1-r_{h}^{4}u^{4}\right)u^{2}}{\left(1-2\alpha r_{h}^{2}u^{2}\right)}\right]^{-1/2}du - \pi,
\label{intTheta}
\end{eqnarray}
where $u_{1} = 1/r_{1}$ with $r_{1}$ being the closest radius to a null geodesic. One way to obtain the classical differential scattering cross section is to analytically solve the integration \eqref{intTheta} by taking the weak-field limit ($r >> r_{h}$) to find
\begin{eqnarray}
\Theta \approx \dfrac{\pi r_{h}^{2}}{16b^{2}(\alpha + 1)^{3/2}}\left[\dfrac{15r_{h}^{2}}{b^{2}(\alpha+1)} - 8\alpha \right].
\label{Theta}
\end{eqnarray}
The classical differential scattering cross section is given in terms of the impact parameter and the scattering angle~\cite{collins1973elastic}
\begin{eqnarray}
\dfrac{d\sigma}{d\Omega} = \sum_{n}\dfrac{b_{n}}{\sin \theta}\Big|\dfrac{db_{n}}{d\theta}\Big|.
\label{scattClassint}
\end{eqnarray}
Here the sum takes into account that $b$ is very close to the critical impact parameter, such that one can orbit the black hole several times before being scattered, in the way that the scattering angle is related to the deflection angle as follows $\Theta = \pm\theta - 2n\pi$, with $n = 0, 1, 2, \ldots$. 
So for our study, we can use the first spiral $n=0$ and obtain the scattering cross section for small angles by combining the equations \eqref{Theta} and \eqref{scattClassint}. Now assuming small $\alpha$ we have
{
\begin{eqnarray}
\dfrac{d\sigma}{d\Omega} \approx \dfrac{r_{h}^{2}\pi}{16 \theta^{3}}\left[\dfrac{\sqrt{15\theta}}{\sqrt{\pi}}\left(1-\dfrac{5\alpha}{4}\right) - 2\alpha \right].
\label{scattClass}
\end{eqnarray}}
We can verify that for $\alpha = 0$, the result agrees with the scattering studies done by Dolan \cite{dolan2009scattering} where the differential scattering cross section for the weak-field deflection angle is proportional to $\theta^{-5/2}$. {For $\alpha \neq 0$ we have the contribution of $\alpha$ in terms of order $\theta^{-5/2}$ and $\theta^{-3}$ causing the scattering cross section to reduce in small $\theta$ regime as we can see in Figure \ref{figScattClas}.}

\subsection{Partial wave method}

\begin{figure}[!htb]
    \centering
    \subfigure[]{\includegraphics[scale=0.40]{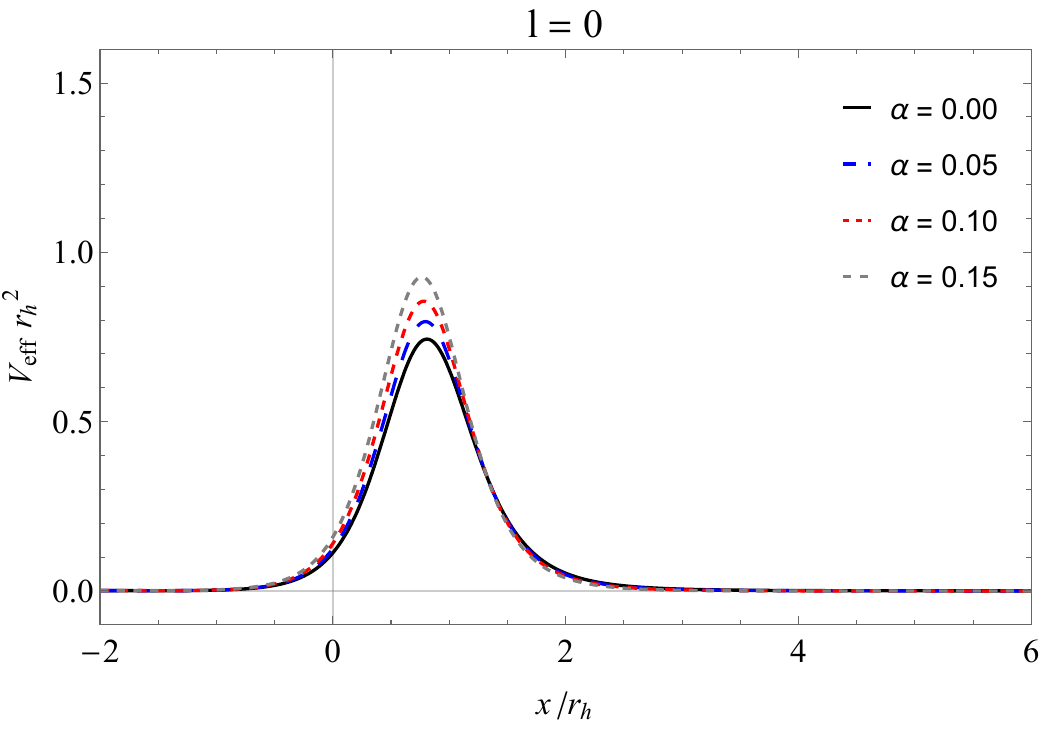}\label{Potl0}}
    \qquad
    \subfigure[]{\includegraphics[scale=0.40]{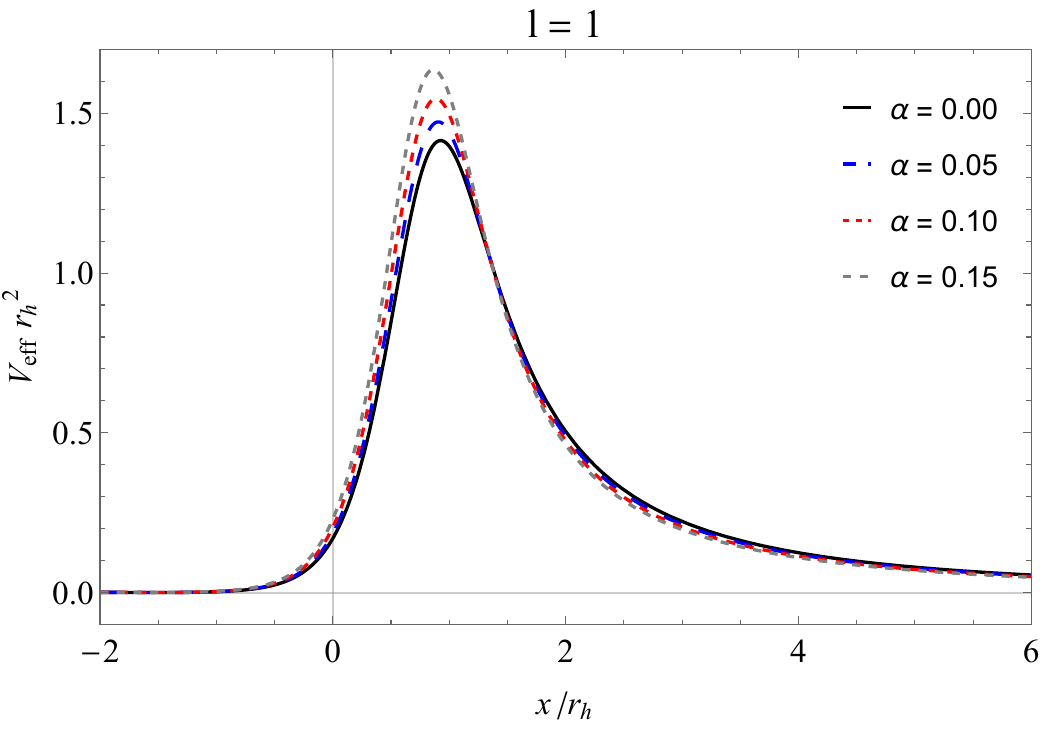}\label{Potl1}}
    \caption{\footnotesize{Effective potential as a function of the tortoise coordinate $x$. We see the effect of the parameter $\alpha$ in the increase of potential.}}
    \label{figpot}
\end{figure}

In this section, we will study the differential scattering cross section and the absorption cross section using the partial wave method in the low energy limit for the metric \eqref{ds}.
We start analyzing the equation for a massless scalar field
\begin{equation}
\frac{1}{\sqrt{-g}}\partial_{\mu}\left(\sqrt{-g}g^{\mu\nu}\partial_{\nu}\Psi \right)=0,
\label{eqKG}
\end{equation}
by first examining the waves scattered across the background. By using the following separation of variables 
\begin{equation}
\Psi_{\omega lm}\left(\vec{r},t\right)=\frac{\mathcal{R}_{\omega l}(r)}{\rho(r)}Y(\theta,\phi)e^{-i\omega t},
\end{equation}
in the Klein-Gordon equation \eqref{eqKG}, 
where $Y(\theta,\phi)$ are the spherical harmonics, then we obtain the radial equation for $\mathcal{R}_{\omega l}$ as follows 
\begin{equation}
\Lambda(r)\frac{d}{dr}\left[\Lambda(r)\frac{d\mathcal{R}_{\omega l}}{dr}\right] + \left(\tilde{\omega}^{2} - V_{eff}\right)\mathcal{R}_{wl}=0,
\label{eqRad1}
\end{equation}
where $\tilde{\omega} = \omega/\left(\alpha + 1\right)$, $\Lambda(r) = \left(1 - \dfrac{r_{h}^{4}}{r^{4}}\right)$ and the effective potential is defined as
\begin{eqnarray}
V_{eff} &=& \dfrac{\Lambda(r)}{\rho(r)}\left( \Lambda'(r) \rho'(r) + \Lambda(r)\rho''(r) + \dfrac{\tilde{l}(\tilde{l}+1)}{\rho(r)}\right), 
\label{poteff}
\end{eqnarray}
being $\tilde{l} = -1/2 +\sqrt{\alpha + 1 +4l(l+1)}/2\sqrt{\alpha +1}$ and the first and second derivatives of $\rho(r)$ given by
\begin{eqnarray}
\rho'(r) = \dfrac{r}{\sqrt{r^{2}-2\alpha r_{h}^{2}}}, \qquad \rho''(r) = \dfrac{-2\alpha r_{h}^{2}}{\left( r^{2}-2\alpha r_{h}^{2}\right)^{3/2}}.
\end{eqnarray}
A very common way to analyze the solution of the radial equation \eqref{eqRad1} in the asymptotic limits, is to introduce a new coordinate $x$ commonly called tortoise coordinate, in the way that the radial equation is transformed into a Schr\"odinger-like equation
\begin{eqnarray}
\dfrac{d^{2}\mathcal{R}_{\omega l}}{dx^{2}} + \left(\tilde{\omega}^{2} - V_{eff}\right)\mathcal{R}_{\omega l} = 0,
\label{eqRSchro}
\end{eqnarray}
where the new coordinate is given by
\begin{eqnarray}
x = r + \dfrac{r_{h}}{4}\log\Big|\dfrac{r-r_{h}}{r+r_{h}}\Big| - \dfrac{r_{h}}{2}\tan^{-1}\left(\dfrac{r}{r_{h}}\right) + \dfrac{\pi r_{h}}{4}.
\end{eqnarray}
The behavior of the effective potential $V_{eff}$ that determines the dynamics of the massless scalar field around the analogous black hole can be seen in the graphs of Figure \ref{figpot}. Notice that even if the potential barrier increases by increasing $\alpha$ and $l$, the potential goes asymptotically to zero, i.e., for $x\rightarrow - \infty \, (r\rightarrow r_{h})$ and $x\rightarrow  \infty \, (r\rightarrow \infty)$. Thus, as the potential goes to zero close to the horizon, we can conclude that for this regime we have a solution of the form $e^{-i\tilde{\omega}x}$. For the case where $r >> r_{h}$ the dominant term in the potential is $\tilde{l}(\tilde{l}+1)/r^{2}$ and the solution can be represented in the form
\begin{eqnarray}
\mathcal{R}_{\omega l} \sim \tilde{\omega}x\left[A_{\omega l}^{out}i^{\tilde{l}+1}h_{\tilde{l}}^{(1)}(\tilde{\omega}x) + A_{\omega l}^{in}(-i)^{\tilde{l}+1}h_{\tilde{l}}^{(1)*}(\tilde{\omega}x) \right],
\label{solRgrande}
\end{eqnarray}
where $h_{\tilde{l}}^{(1)}(\tilde{\omega} x)$ is the spherical Bessel functions of the third kind~\cite{abramowitz1965handbook}, $A_{\omega l}^{in}$ and $A_{\omega l}^{out}$ are complex constants. Remembering that $h_{\tilde{l}}^{(1)}(\tilde{\omega} x) \simeq (-i)^{\tilde{l}+1}e^{i\tilde{\omega}x}/(\tilde{\omega} x)$ when $\tilde{\omega} x >> \tilde{l}(\tilde{l}+1)/2$, we have the following solutions
\begin{eqnarray}
   \mathcal{R}_{\omega l} \approx  A_{\omega l}^{out}e^{i\tilde{\omega}x} + A_{\omega l}^{in}e^{-i\tilde{\omega}x}, \qquad (x \rightarrow  + \infty).
   \label{condAssintotic1}
\end{eqnarray}
Now, as the potential goes to zero at the event horizon, we have solution of the form
\begin{eqnarray}
   \mathcal{R}_{\omega l} \approx  A_{\omega l}^{tr}e^{-i\tilde{\omega}x}, \qquad (x \rightarrow - \infty).
   \label{condAssintotic2}
\end{eqnarray}
The phase shift is obtained by the relation
\begin{eqnarray}
e^{2i\delta_{l}} = (-1)^{l+1}\dfrac{A_{\omega l}^{out}}{A_{\omega l}^{in}}.
\label{dfase}
\end{eqnarray}
 
Now, we write the equation \eqref{eqRad1} in Schr\"odinger-like equation form using the following change of variable in the radial function $X(r)= \sqrt{\Lambda(r)}\mathcal{R}$
\begin{eqnarray}
\frac{d^{2}X(r)}{dr^{2}} + U(r)X(r) = 0,
\label{eqRadSchro}
\end{eqnarray}
with the potential
\begin{eqnarray}
U(r) = \dfrac{1}{\Lambda(r)^{2}}\left[\tilde{\omega}^{2}+ \frac{\left(\Lambda^{\prime}(r)\right)^{2}}{4} - \frac{\Lambda(r)\Lambda^{\prime\prime}(r)}{2} - V_{eff}\right].
\label{pot}
\end{eqnarray}
We expand the potential \eqref{pot} in terms of $1/r$, and we obtain a new expression for the radial equation
\begin{equation}
\frac{d^{2}X(r)}{dr^{2}} + \left[\tilde{\omega}^{2} - \dfrac{\tilde{l}(\tilde{l}+1)}{r^{2}} + \mathcal{U}_{eff}(r)\right]X(r) = 0.
\label{eqRadexp}
\end{equation}
We can define a new effective potential of the form
\begin{eqnarray}
\mathcal{U}_{eff}(r) &=&\dfrac{2r_{h}^{2}\tilde{\omega}^{2}\left[r^2_h\tilde{\omega}^{2} - \alpha\tilde{l}(\tilde{l}+1) + \alpha\right]}{\tilde{\omega}^{2}r^{4}} - \dfrac{r^4_h\tilde{\omega}^{4} \left[\tilde{l}(\tilde{l}+1)(1+4\alpha^{2}) - 6 - 8\alpha^{2}\right]}{\tilde{\omega}^{4}r^{6}} \nonumber \\
 & & +\dfrac{3r_{h}^{8}\tilde{\omega}^{8} - 2r_{h}^{6} \alpha\left[4 + \tilde{l}(\tilde{l}+1)(1+4\alpha^{2}) - 12\alpha^{2}\right]\tilde{\omega}^{6}}{\tilde{\omega}^{6}r^{8}}+ \cdots + \mathcal{O}\left(1/\tilde{\omega}^{8}r^{10}\right).
 \label{potExpand}
\end{eqnarray}
At this point, to obtain the differential scattering cross section in the large $l$ limit, we apply the Born approximation~\cite{morse1954methods} to determine the phase shift $\delta_l$, as follows
\begin{eqnarray}
\delta_{l} \approx -\tilde{\omega}\int_{0}^{\infty}r^{2}J_{\tilde{l}}^{2}(\tilde{\omega} r)\mathcal{U}_{eff}(r)dr.
\end{eqnarray}
We can use the first two terms of the potential expansion $\mathcal{U}_{eff}(r)$ for a contribution up to the order 
${\cal O}[(\tilde{\omega}r_{h})^{4}]$. Now applying the following identities
\begin{eqnarray}
\int_{0}^{\infty} z^{-2}J_{l}^{2}(z)dz &=& \dfrac{\pi}{(2l-1)(2l+1)(2l+3)},\\
\int_{0}^{\infty} z^{-4}J_{l}^{2}(z)dz &=& \dfrac{3\pi}{(2l-3)(2l-1)(2l+1)(2l+3)(2l+5)},
\end{eqnarray}
we can obtain the phase shift given as follows 
\begin{eqnarray}
\delta_{\tilde{l}} \approx -\dfrac{2\pi r_{h}^{2}\tilde{\omega}^{2}\left[r_{h}^{2}\tilde{\omega}^{2} - \alpha\left(\tilde{l}(\tilde{l}+1)-1\right)\right]}{(2\tilde{l} - 1)(2\tilde{l} + 1)(2\tilde{l} + 3)} + \dfrac{3\pi r_{h}^{4}\tilde{\omega}^{4}\left[\tilde{l}(\tilde{l}+1)(1+4\alpha^{2}) -6 -8\alpha^{2}\right]}{(2\tilde{l} -3)(2\tilde{l} -1)(2\tilde{l} + 1)(2\tilde{l} + 3)(2\tilde{l} + 5)}.
\end{eqnarray}
Applying the limit of large $l$ in the equation just below \eqref{poteff}, we have $\tilde{l} \approx l/\sqrt{1+\alpha}$. Thus the equation for the approximate phase shift in terms of $(l + 1/2)$ is given in the form 
\begin{eqnarray}
\delta_{l} \approx \dfrac{\alpha \pi r_{h}^{2}\omega^{2}}{4\left(l+1/2\right)(1+\alpha)^{3/2}} - \dfrac{5 \pi r_{h}^{4}\omega^{4}}{32\left(l+1/2\right)^{3}(1+\alpha)^{5/2}}.
\end{eqnarray}
This result can be compared with the semi-classical scattering section \eqref{Theta} in the large $l$ limit. Through the semi-classical description of scattering by Ford and Wheeler \cite{ford1959semiclassical,Ford:2000uye}, where the impact parameter is associated with each partial wave $b = (l+1/2)/\omega$, using the deflection function with the weak-field deflection angle, we find
\begin{eqnarray}
\Theta = \dfrac{d}{dl}\left[ Re(2\delta_{l})\right] &=& \dfrac{15r_{h}^{4}\omega^{4}\pi}{16(\alpha + 1)^{5/2}(1+1/2)^{4}} - \dfrac{r_{h}^{2}\omega^{2}\alpha\pi}{2(\alpha + 1)^{3/2}\left(l+1/2\right)^2}, \nonumber\\
&=& \dfrac{r_{h}^{2}\omega^{2}\pi}{16b^{2}(\alpha+1)^{3/2}}\left[\dfrac{15}{(\alpha + 1)b^{2}} - 8\alpha\right].
\end{eqnarray}
Notice that the results precisely match those obtained in the equation \eqref{Theta}.

 \subsection{Absorption cross section at low frequencies}
 Here, to find the absorption cross section at the low-frequency limit, we will determine the phase shift $\delta_l$ at two different  approximate approaches. Firstly, we will solve the radial equation and secondly, we will analyze the effective potential expanded in powers of $1/r$, to determine  $\delta_l$.
 \subsubsection{Phase Shift: Solution of the Radial Equation}
\begin{figure}[!htb]
    \centering
    \subfigure[]{\includegraphics[scale=0.34]{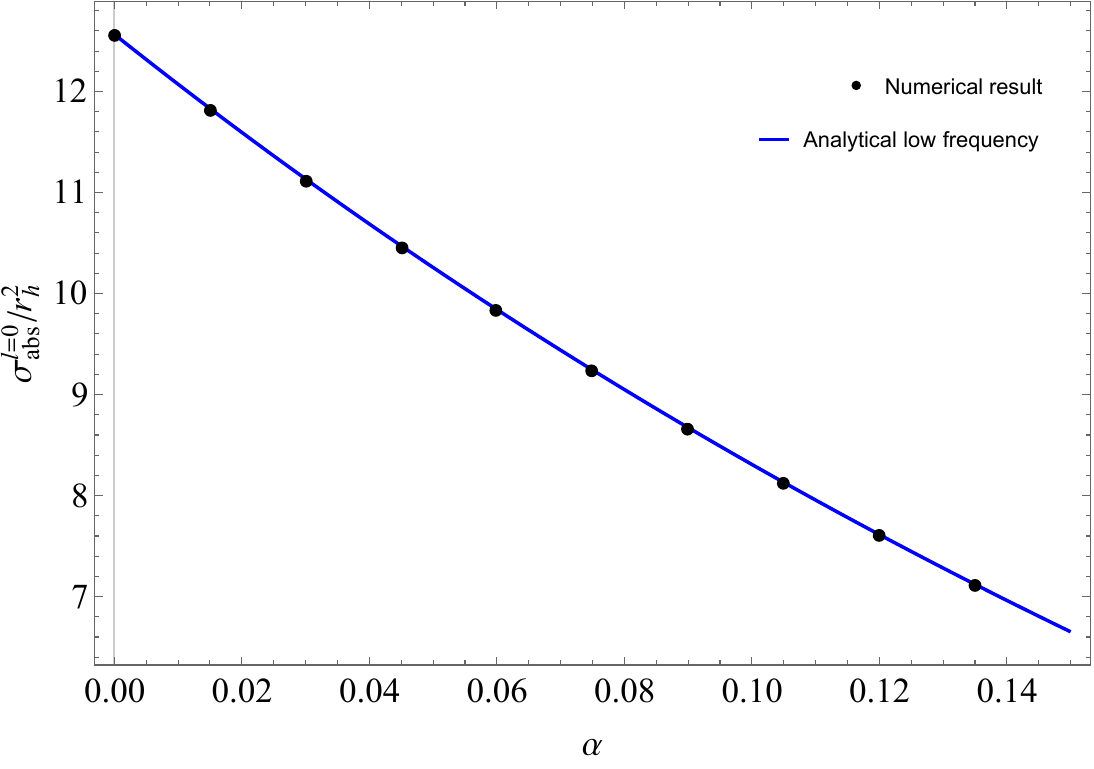}\label{abslowfre}}
    \quad
    \subfigure[]{\includegraphics[scale=0.30]{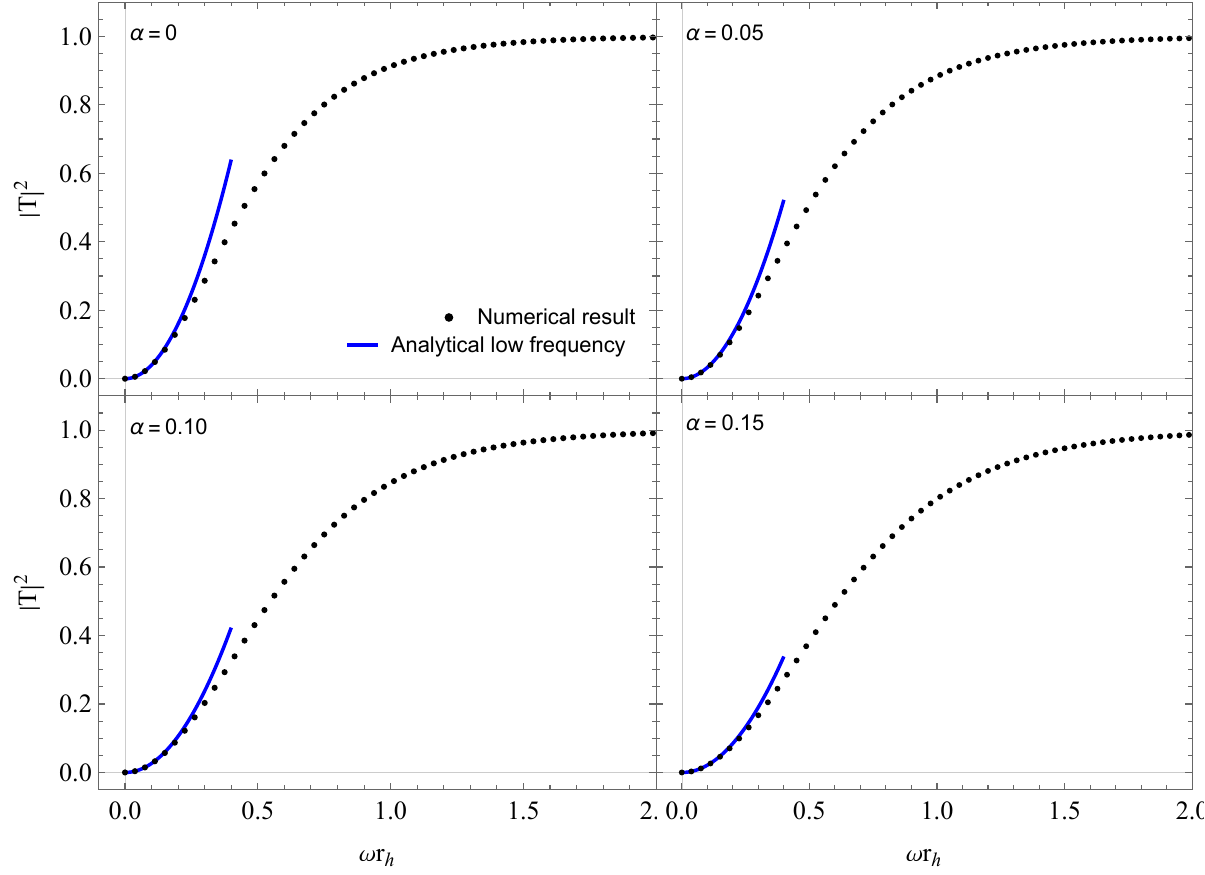}\label{TrasLF}}
    \caption{\footnotesize{In graph (a) we have absorption cross section at low frequencies as a function of the parameter $\alpha$. In graph (b) we have a comparison between the numerical transmission coefficient with that obtained analytically at low frequencies.}}
    \label{AbsTrans}
\end{figure}

The absorption cross section at low frequencies can be obtained by approximating the radial equation \eqref{eqRad1} at asymptotic limits assuming the condition of $\omega \rightarrow 0$. In \cite{arfken1999mathematical},  the absorption cross section at low frequencies is analyzed in a generalized way for spherically symmetric black holes for a massless scalar field. It was found that the cross section is proportional to the area of the black hole.
For our case, we will find a solution to the radial equation \eqref{eqRad1} in the low frequency regime
\begin{eqnarray}
   r(r^{4}-r_{h}^{4})\frac{d^{2}\mathcal{R}_{\omega \tilde{l}}}{dr^{2}} + 4r_{h}^{4}\frac{d\mathcal{R}_{\omega \tilde{l}}}{dr} - \dfrac{r}{(r^{2}-2\alpha r_{h}^{2})}\left[4r_{h}^{4} - \dfrac{2\alpha r_{h}^{2}(r^{4}-r_{h}^{4})}{(r^{2}-2\alpha r_{h}^{2})} + \tilde{l}(\tilde{l}+1)r^{4}\right]\mathcal{R}_{\omega \tilde{l}}=0.
\end{eqnarray}
 For this, we can assume the following change of variable $\rho = \sqrt{r^{2} - 2\alpha r_{h}^{2}}$
\begin{eqnarray}
	\left((\rho^{2} + 2\alpha r_{h}^{2})^{2} - r_{h}^{4}\right)(\rho^{2} + 2\alpha r_{h}^{2})\dfrac{d^{2}\mathcal{R}_{\omega \tilde{l}}}{d\rho^{2}} + \left[4r_{h}^{4}\rho - \dfrac{2\alpha r_{h}^{2}}{\rho}\left((\rho^{2} + 2\alpha r_{h}^{2})^{2} - r_{h}^{4}\right) \right]\dfrac{d\mathcal{R}_{\omega \tilde{l}}}{d\rho} -  \nonumber \\ 
	\left[4r_{h}^{4} - \dfrac{2\alpha r_{h}^{2}\left((\rho^{2} + 2\alpha r_{h}^{2})^{2} - r_{h}^{4}\right)}{\rho^{2}} 
	 + \tilde{l}(\tilde{l}+1)\left(\rho^{2} + 2\alpha r_{h}^{2}\right)^{2}\right]\mathcal{R}_{\omega \tilde{l}} = 0.
\end{eqnarray}
At spatial infinity (large $\rho$ limit) the radial equation is of the form
\begin{eqnarray}
	\rho^{4} \dfrac{d^{2}\mathcal{R}_{\omega \tilde{l}}}{d\rho^{2}}  - 2\alpha \rho r_{h}^{2}\dfrac{d\mathcal{R}_{\omega \tilde{l}}}{d\rho} -   
	\left[ - 2\alpha r_{h}^{2} + \tilde{l}(\tilde{l}+1)\rho^{2} \right]\mathcal{R}_{\omega \tilde{l}} = 0.
\end{eqnarray}
The solution of the above equation is given in terms of hypergeometric functions as follows
\begin{eqnarray}
	\mathcal{R}_{\omega \tilde{l}} = C_{1} \left(\dfrac{\rho}{\sqrt{\alpha} r_{h}}\right)^{(1+\tilde{l})}{_{1}F_{1}}\left(-\dfrac{\tilde{l}}{2}, \dfrac{1}{2} - \tilde{l},-\dfrac{r_{h}^{2}\alpha}{\rho^{2}}\right)
	+ C_{2} \left(\dfrac{r_{h}\sqrt{\alpha}}{\rho}\right)^{\tilde{l}}{_{1}F_{1}}\left(\dfrac{1+\tilde{l}}{2}, \dfrac{3}{2} + \tilde{l},-\dfrac{r_{h}^{2}\alpha}{\rho^{2}}\right),
\end{eqnarray}
where $_{1}F_{1}$ is a hypergeometric functions, $C_{1}$ and $C_{2}$ are constants to be determined. For the absorption cross section at low frequencies, the greatest contribution comes from the partial waves, where $\tilde{l} = 0 (l=0)$. By using this and the fact that we are in the regime of $\rho \rightarrow \infty $, the solution becomes
	\begin{eqnarray}
	\mathcal{R}_{\omega 0} = \dfrac{\rho  C_{1}}{\sqrt{\alpha} r_{h}}
	+  \left(1 - \dfrac{r_{h}^{2}\alpha}{3 \rho^{2}}\right)C_{2}.
	\label{solR00}
	\end{eqnarray}
Now we need to find the constants $C_{1}$ and $C_{2}$. In order for that we rewrite the solution \eqref{solRgrande} at low frequencies $\tilde{\omega} x <<1$, by using the relation for the Bessel function~\cite{arfken1999mathematical}
\begin{eqnarray}
 h_{\tilde{l}}^{(1)} \approx \dfrac{2^{\tilde{l}}\tilde{l}!}{(2\tilde{l}+1)!}(\tilde{\omega}x)^{\tilde{l}} - i\dfrac{(2\tilde{l})!}{2^{\tilde{l}}\tilde{l}!}(\tilde{\omega}x)^{-\tilde{l}-1},\qquad 
 (\tilde{\omega} x <<1 ).   
\end{eqnarray}
We now find
\begin{eqnarray}
\mathcal{R}_{\omega \tilde{l}} \sim \left[(-i)^{\tilde{l}+1}A_{\omega l}^{in} + i^{\tilde{l}+1}A_{\omega l}^{out}\right]\dfrac{2^{\tilde{l}}\tilde{l}!}{(2\tilde{l} + 1)!}(\tilde{\omega}x)^{\tilde{l}+1} + i\left[(-i)^{\tilde{l}+1}A_{\omega l}^{in} - i^{\tilde{l}+1}A_{\omega l}^{out}\right]\dfrac{(2\tilde{l})!}{2^{\tilde{l}}\tilde{l}!}(\tilde{\omega}x)^{-\tilde{l}},
\label{solRlf}
\end{eqnarray}
and in order for to guarantee that the solution at low frequencies is not divergent, 
{the second term of equation \eqref{solRlf} has to satisfy the relation $A_{\omega l}^{in} \approx (-1)^{\tilde{l}+1}A_{\omega l}^{out}+ \cdots +\mathcal{O}(\omega)$. In this way the wave is almost completely spread out in the low frequency regime, and the solution reads}
\begin{eqnarray}
\mathcal{R}_{\omega \tilde{l}} \sim (-i)^{\tilde{l}+1}A_{\omega l}^{in}\dfrac{2^{\tilde{l}+1}\tilde{l}!}{(2\tilde{l} + 1)!}(\tilde{\omega}x)^{\tilde{l}+1}.
\end{eqnarray}
For $\tilde{l}=0$, we have $\mathcal{R}_{\omega 0} = -2i(\tilde{\omega} x)A_{\omega l}^{in}$, and by 
comparing with the solution \eqref{solR00}, we conclude that $C_{2} = 0$ and $C_{1} = -2i\sqrt{\alpha}A_{\omega l}^{in}\tilde{\omega}r_{h}$. Then, the solution \eqref{solR00} is of the form
\begin{eqnarray}
\mathcal{R}_{\omega 0} = -2iA_{\omega l}^{in}\tilde{\omega}\rho.
\label{sol_inf}
\end{eqnarray}
Now, by using the equation \eqref{condAssintotic2} in the low frequency limit $(\tilde{\omega}x << 1)$, the expression can be approximated as $\mathcal{R}_{\omega l} \approx A_{\omega l}^{tr}$. This result can be combined with \eqref{sol_inf} by applying the limit $\rho\rightarrow r_{h}\sqrt{1-2\alpha}$ to find the following relation $-2iA_{\omega l}^{in}\tilde{\omega}r_{h}\sqrt{1-2\alpha} = A_{\omega l}^{tr}$ and then we can obtain the transmission coefficient
\begin{eqnarray}
\dfrac{A_{\omega l}^{tr}}{A_{\omega l}^{in}} = -2i\tilde{\omega}r_{h}\sqrt{1-2\alpha},\quad \Rightarrow \quad |T|^{2} = \Big|\dfrac{A_{\omega l}^{tr}}{A_{\omega l}^{in}}\Big|^{2} = 4\tilde{\omega}^{2}r_{h}^{2}(1-2\alpha).
\label{eqTrans}
\end{eqnarray}
The absorption cross section in terms of the transmission coefficient $T$ is given by
\begin{eqnarray}
\sigma_{abs} &=& \dfrac{\pi}{\omega^{2}}\sum_{l=0}(2l+1)|T|^{2},\nonumber \\
\sigma_{abs}^{(0)} &=& \dfrac{4\pi r_{h}^{2}(1 - 2\alpha)}{(1+\alpha)^{2}} =  \dfrac{4\pi \rho^{2}_h}{(\alpha + 1)^{2}}.
\label{abslf}
\end{eqnarray}
Note that the result for the absorption cross section at low frequencies is not directly proportional to a horizon radius area as in the usual canonical case \cite{crispino2007absorption}, but rather is proportional to a new area of radius $\rho(r_{h})$ due to the term introduced by the symmetry breaking.
In Figure \ref{abslowfre}, we compare the numerical result with those obtained in equation \eqref{abslf}. We can see that the points obtained numerically fit the blue curve, which is the cross section of absorption as a function of the parameter $\alpha$. We can also compare the transmission coefficient $|T|$ obtained above with the one obtained numerically, as we can see in Figure \ref{TrasLF}. {The blue curves represent the results obtained from \eqref{eqTrans} for different values of $\alpha$ with $0\leq \omega \leq 0.4$, we can see that for low frequencies the analytical and numerical results (dotted line) fit.} 
\subsubsection{Phase Shift: Effective Potential}
Next, we will determine the phase shift $\delta_l$ by analyzing the effective potential (\ref{potExpand}) following the method employed in Refs. \cite{Anacleto:2017kmg,anacleto2020absorption,Anacleto:2020zhp,Anacleto:2020lel,Anacleto:2022shk}.
In the gravitational case, for example the Schwarzschild black hole, where we have the term $r_h/r$ in the metric function, we determine the $\delta_l$ by analyzing the term $1/r^2$ in the effective potential. In the case of the canonical acoustic black hole, we have the contribution of $r^4_h/r^4$ in the metric function, and then the phase shift will be determined from the term $1/r^6 $ that appears in the effective potential after performing the expansion in $1/r$.
For our case, we use the coefficient referring to the term $1/r^{6}$ written in terms of the suitable horizon radius, $\rho_h$, as follows:
\begin{eqnarray}
 \dfrac{\tilde{\omega}^{4}r_{h}^{4}\left[6 -\tilde{l}(\tilde{l}+1)\left(1+4\alpha^{2}\right) + 8\alpha^{2}\right]}{\tilde{\omega}^{4}r^{6}}
 =\dfrac{6\tilde{\omega}^{4}(1+4\alpha)\rho_{h}^{4}\left[1 -\tilde{l}(\tilde{l}+1)/6\right]}{\tilde{\omega}^{4}r^{6}}  + O(\alpha^2) = \dfrac{6(1+4\alpha)\tilde{\omega}^{2}\rho_{h}^{2}\ell^{2}}{\tilde{\omega}^{2}r^{6}}  + O(\tilde{\alpha}^2)
\end{eqnarray}
where we define
\begin{eqnarray}
 \ell^2 = \rho_{h}^{2} \tilde{\omega}^{2}\left[1 - \frac{l(l+1)}{6(1+\alpha)}\right].   
\end{eqnarray}
Then, applying the approximation formula, we have
\begin{eqnarray}
 \delta_l \approx l - \ell
 =l -\tilde{\omega}\rho_{h}\left[1 -\dfrac{l(l+1)}{6(1+\alpha)} \right]^{1/2}.
 \label{dfase2}
\end{eqnarray}
Thus, in the limit $l\rightarrow 0$ we find the phase shift given by
\begin{eqnarray}
  \delta_0 =-\frac{\omega \rho_{h}}{(1+\alpha)}=-\frac{\omega r_{h}\sqrt{1-2\alpha}}{(1+\alpha)}.
  \label{ps0} 
\end{eqnarray}
As obtained from quantum mechanics, the total absorption cross section can be calculated using the following relationship
\begin{eqnarray}
\sigma_{abs} &=& \dfrac{\pi}{\omega^{2}}\sum_{l=0}^{\infty}(2l + 1)\left(1 - |e^{2i\delta_{l}}|^{2}\right) = \dfrac{4\pi}{\omega^{2}}\sum_{l=0}^{\infty}(2l + 1)\sin^{2}(\delta_{l}),\\
&=& \dfrac{4\pi}{\omega^{2}}\left[\sin^{2}(\delta_{0}) + \sum_{l=0}^{\infty}(2l + 1)\sin^{2}(\delta_{l\geq1}) \right].
\end{eqnarray}
Now, by taking low frequency limit with $\delta_{l}$ given by \eqref{ps0} the absorption reads
\begin{eqnarray}
\sigma^{\mathrm{l f}}_{abs}= \dfrac{4\pi \delta^2_{0}}{\omega^{2}}= 4\pi r^2_h \dfrac{\left(1-2\alpha\right)}{\left(1 + \alpha \right)^{2}}=\dfrac{4\pi \rho^2_h}{\left(1 + \alpha \right)^{2}}.
\end{eqnarray}
Therefore, we have obtained the same result as found in~\eqref{abslf}.
	\subsection{Results for absorption and scattering}
Another way to obtain the solution of the radial equation \eqref{eqRad1} and study the effects of the absorption and differential scattering cross sections is solving numerically, by using the asymptotic conditions to obtain the values of the input and output coefficients ($A_{in}$, $A_{out}$) and consequently the phase shift using the relation \eqref{dfase}. The obtained results cover the entire frequency spectrum for the absorption cross section with differential scattering at larger angles.

\begin{figure}[!htb]
    \centering
    \subfigure[]{\includegraphics[scale=0.32]{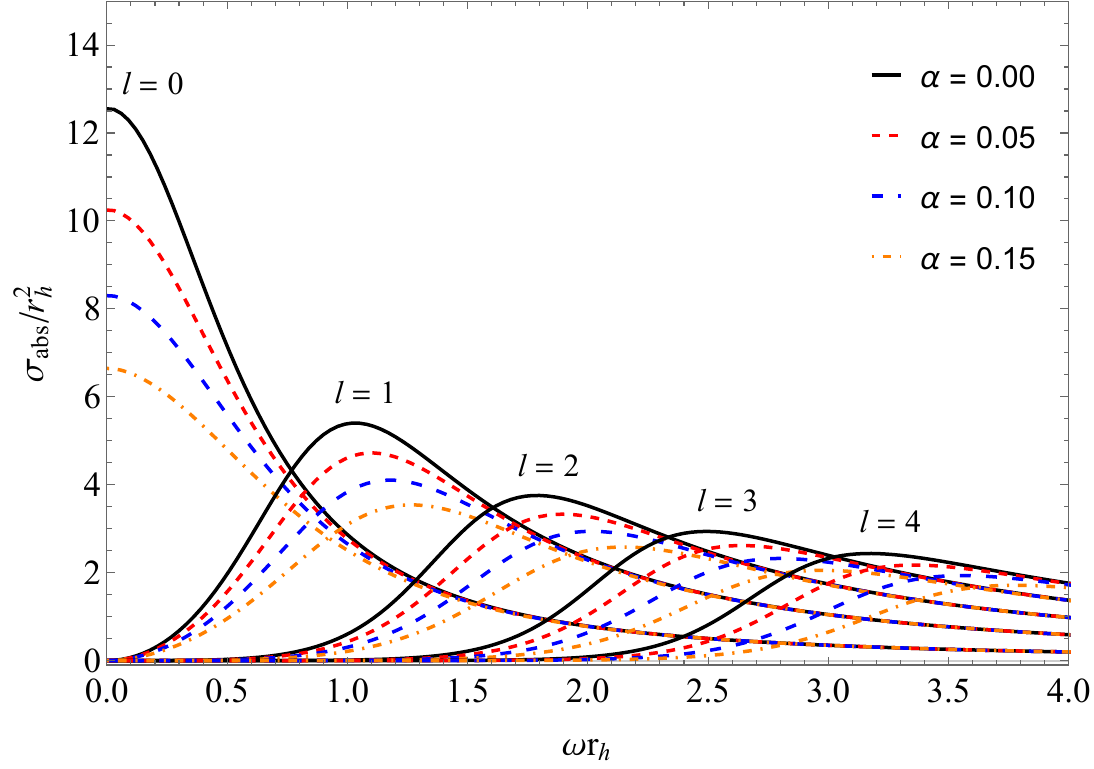}\label{SecaoAbs}}
    \qquad
    \subfigure[]{\includegraphics[scale=0.35]{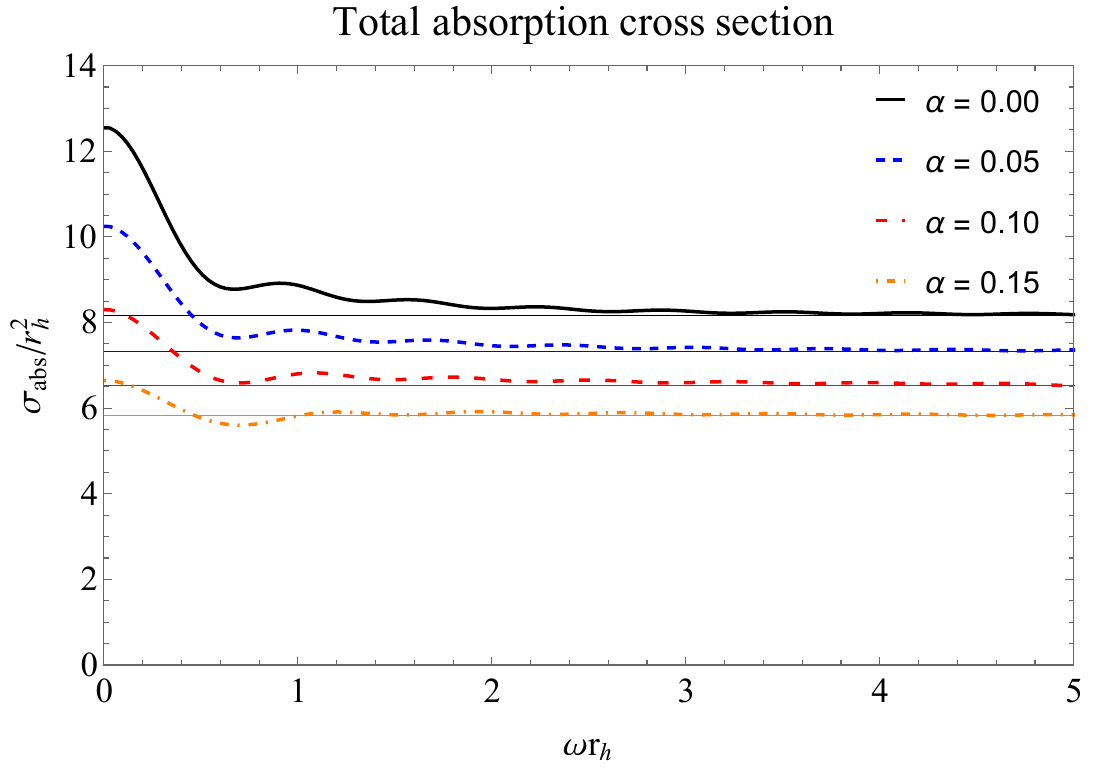}\label{SecaoAbsTotal}}
    \caption{\footnotesize{Absorption cross as a function of frequency. In \ref{SecaoAbs} we have the partial behavior for the multipoles $l=0, 1, 2, 3$, while in \ref{SecaoAbsTotal} we have the total absorption section converging to the lines that represent the results for absorption at high frequencies using the geodesic method.}}
    \label{abs}
\end{figure}

\begin{figure}[!htb]
    \centering
    \subfigure[]{\includegraphics[scale=0.35]{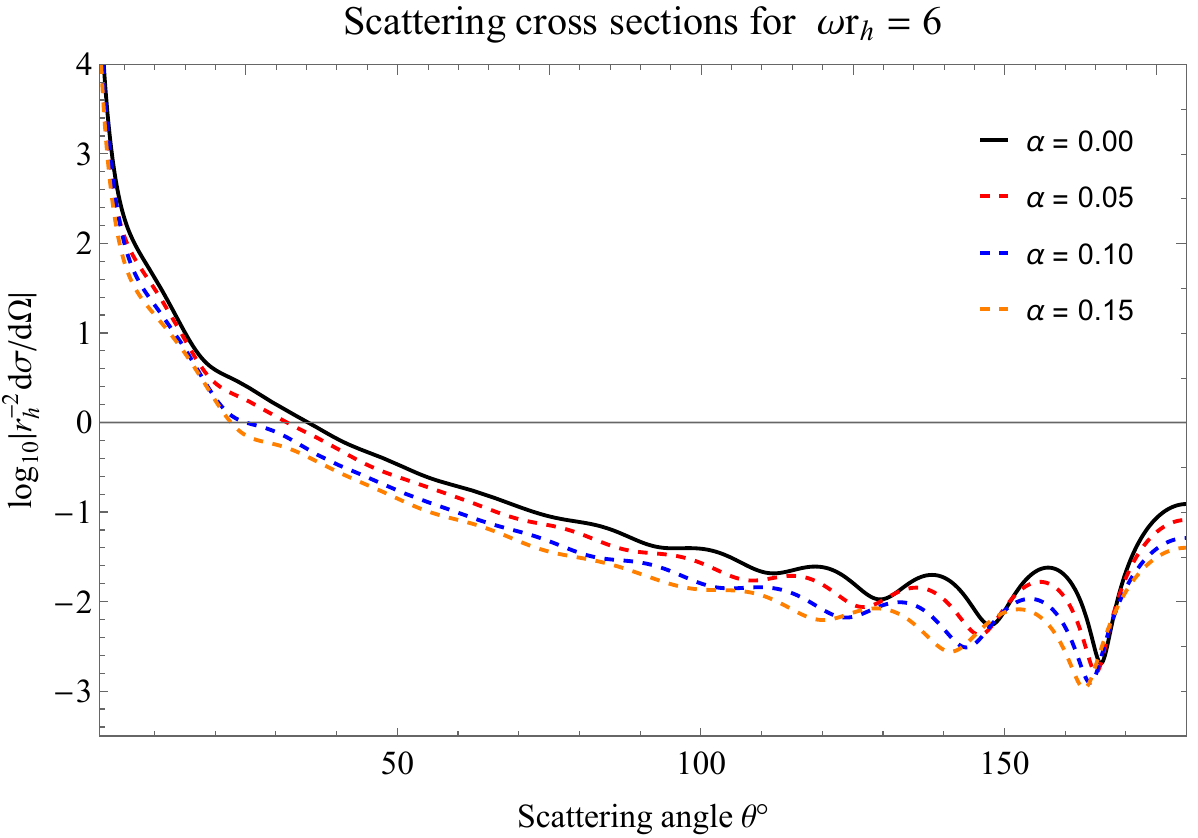}\label{scattw6}}
    \qquad
    \subfigure[]{\includegraphics[scale=0.35]{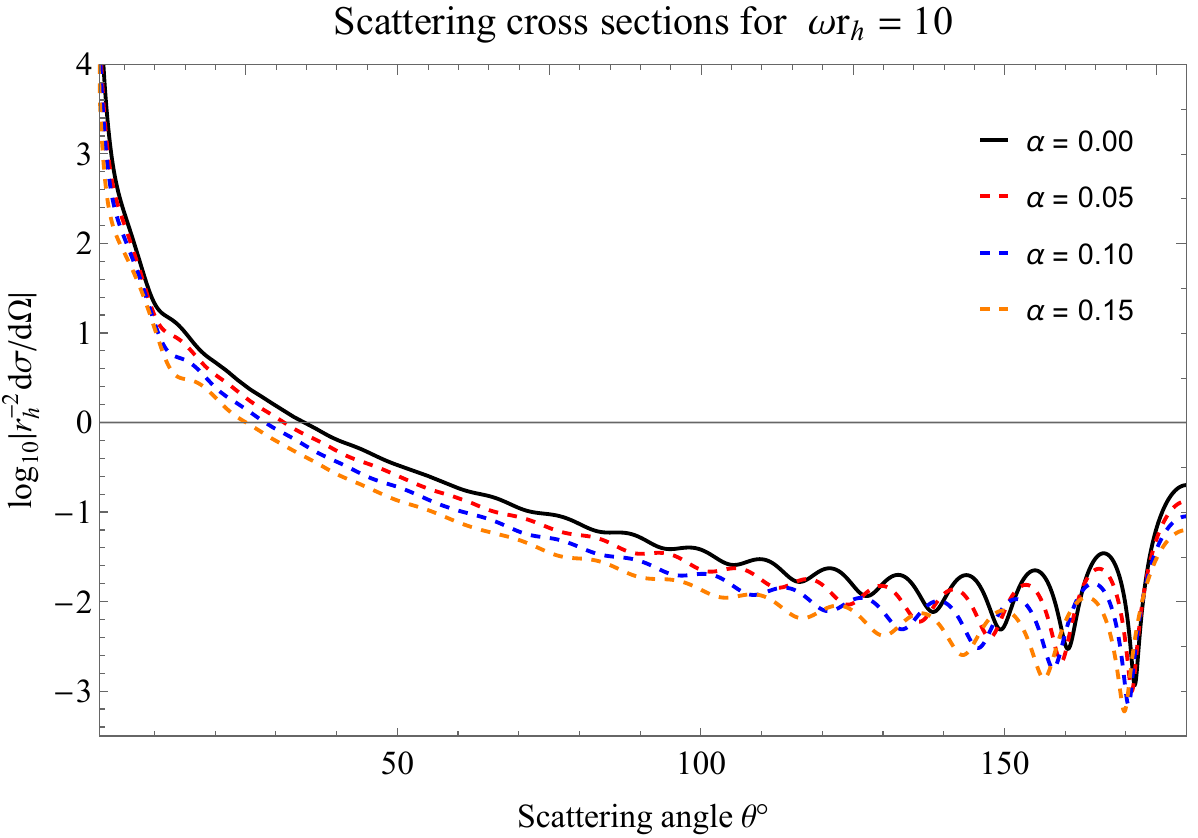}\label{scattw10}}
    \caption{\footnotesize{Numerically obtained differential scattering cross section for two different frequency levels.}}
    \label{scatt}
\end{figure}

\begin{figure}[!htb]
 \centering
 \includegraphics[scale=0.35]{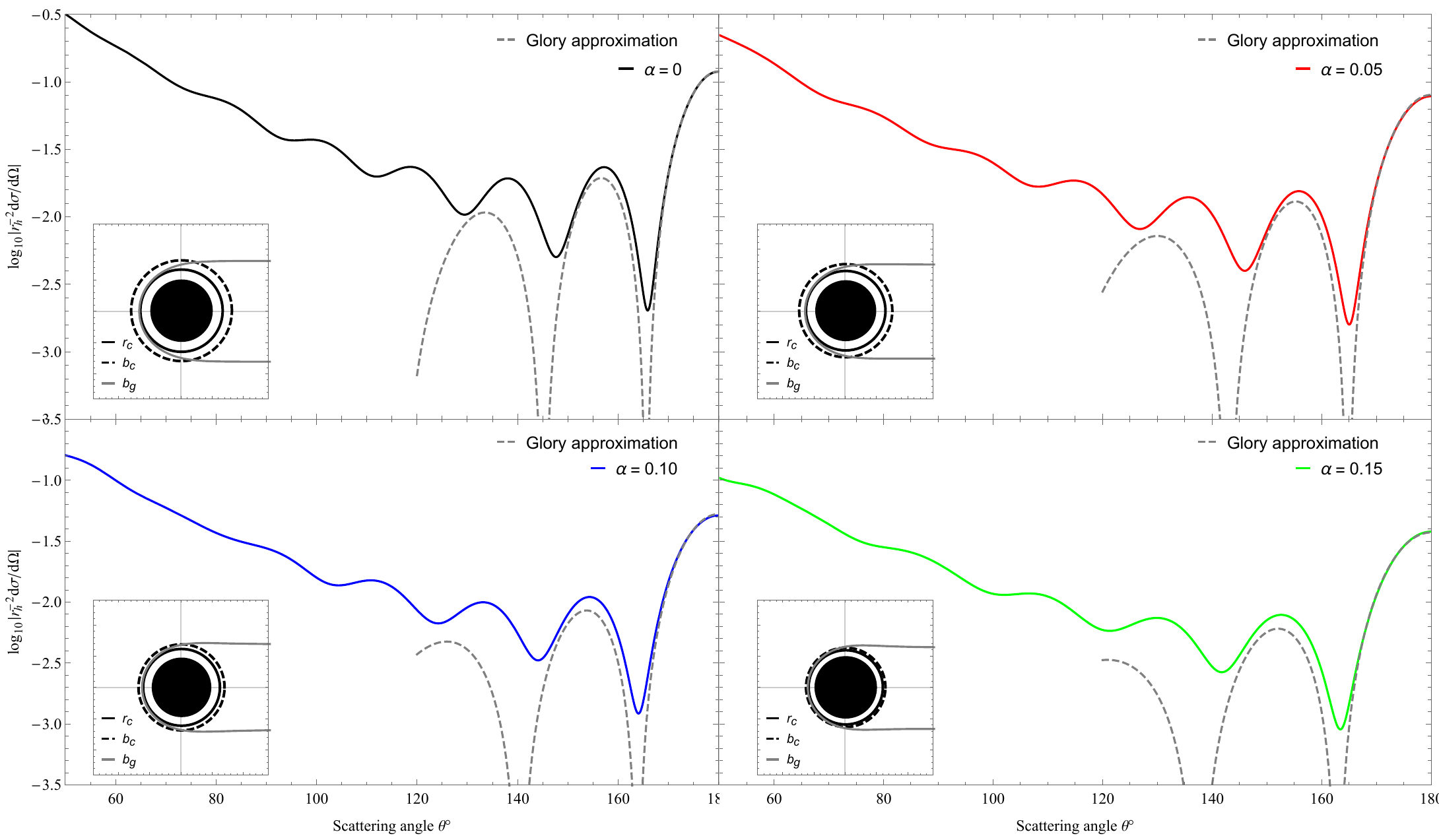}
  \caption{\footnotesize{Comparison between differential scattering cross section partial wave method and the glory approximation. The inset in each graph shows the behavior of the geodesic line for $b_{g}$ found.}}
 \label{figGlory}
\end{figure}

In Figure \ref{abs}, we can see the partial and total absorption cross section as a function of frequency. The influence of the parameter $\alpha$ reduces the the absorption cross section.
In Figure \ref{scatt}, we have the behavior for the differential scattering cross section for two frequency spectra.
{The numerical results show that the $\alpha$ parameter influences the differential scattering cross sections very few at small angles, in agreement with the analytical studies for the classical case obtained in \eqref{scattClass}. For large angles, the contribution from Lorentz symmetry breaking causes a shift in the interference fringes.}
We can complement the scattering study and verify the results obtained in Fig.~\ref{scatt} at larger angles close to $\pi$. For this, we use the so-called semi-classical glory approximation. For a spherically symmetric black hole, we have the following formula~\cite{matzner1985glory}
\begin{eqnarray}
\dfrac{d\sigma}{d\Omega}\Big |_{\theta \approx \pi} \approx 2\pi \omega b_{g}^{2} \Big | \dfrac{d b}{d \theta}  \Big |_{\theta = \pi}\left[ J_{2s}\left(\omega b_{g}\sin\theta \right)\right]^{2},
\label{GloryScatteq}
\end{eqnarray}
where $b_{g}$ is the impact parameter for backscattered rays $(\theta = \pi)$, $J_{2s}(x)$ is the Bessel function of the first kind and $s$ is the spin of the field (for the scalar wave $s=0$). 
The value of $b_{g}$ can be obtained numerically from equation \eqref{eqD1}, given a deflection angle of the form
\begin{eqnarray}
\Theta(b) = \dfrac{2}{\sqrt{v_{3}(v_{2}-v_{1})}}\left[K(\kappa) +  \dfrac{2\alpha r_{h}^{2}v_{1}}{1-2\alpha r_{h}^{2}v_{1}} \Pi (p,\kappa)\right] - \pi,
\end{eqnarray}
where $K(\kappa)$ and $\Pi(p,\kappa)$ are complete elliptic integrals of first and third order respectively\cite{slater1965handbook}, with
\begin{eqnarray}
\kappa^{2} = \dfrac{v_{2}(v_{3} - v_{1})}{v_{3}(v_{2} - v_{1})} \qquad \text{and} \qquad p^{2} = \dfrac{v_{3} - v_{1}}{v_{3} - 2\alpha r_{h}^{2} v_{1}v_{3}}.
\end{eqnarray}
 Here, $v_{1}$, $v_{2}$, and $v_{3}$ are the roots of the equation $(1-2\alpha r_{h}^{2} v)/b^{2} - (1+\alpha)(1-r_{h}^{4}v^{2})v = 0$ where we made $u^{2} = v$ in equation \eqref{eqD1}. For geodesic scattering, the roots obey the following relations $v_{1}<0$, $v_{2}>0$ and $v_{3}>v_{2}$.
We now write the differential scattering cross section \eqref{GloryScatteq} as follows
\begin{eqnarray}
\dfrac{1}{r_{h}^{2}}\dfrac{d\sigma}{d\Omega}\Big |_{\theta \approx \pi} \approx \bar{A}\omega r_{h}\left[J_{0}(b_{g}\omega \sin\theta) \right]^{2},
\label{GloryScatteq2}
\end{eqnarray}
where $\bar{A} = \dfrac{2\pi b_{g}^{2}}{r_{h}^{3}}\Big | \dfrac{d b}{d \theta} \Big |$. 
In Table \ref{tab2}, we have some numerical results for the parameters $b_{c}$, $b_{g}$ and $\bar{A}$. We have a reduction in glory peak due to the Lorentz symmetry breaking term. For example, comparing $\alpha=0$ with $\alpha=0.15$, it is approximately $3.2$ times weaker in magnitude compared to the case without the presence of the break of symmetry.
\begin{table}[!ht]
	\begin{center}
	\caption{\footnotesize{Parameters of the glory approximation obtained numerically and critical impact parameter.}} 
	\label{tab2}
		\begin{tabular}{|c||c|c|c|}
	\hline
 $\alpha$ & $b_{c}/r_{h}$ &  $b_{g}/r_{h}$ & $\bar{A}$ \\
 \hline
 0.00  & 1.61185  & 1.61346 &  0.19930$\times 10^{-1}$  \\
 0.05  & 1.52645  & 1.52690 &  0.13351$\times 10^{-1}$  \\
 0.10  & 1.44336  & 1.44369 &  0.87779$\times 10^{-2}$  \\
 0.15  & 1.36178  & 1.36204 &  0.62170$\times 10^{-2}$  \\
  \hline
 		\end{tabular}
	\end{center}
\end{table}
In Fig.~\ref{figGlory}, we compare the results of the glory approximation \eqref{GloryScatteq2} with the results obtained by the partial wave method. Then, we verify the behavior considering a frequency $\omega r_{h} = 6$,  for four values of $\alpha$. Besides, we have for each case an inset showing the backscattered behavior of the geodesic line whose result is obtained by inserting the parameter value of impact $b_{g}$ on the orbit equations \eqref{eqD1} and \eqref{eqD2}. In this case, the differential scattering cross section presents a good approximation for angles close to $\pi$. The oscillations for the case $\alpha =0$ are narrower in relation to the other cases with $\alpha \neq 0$. That is, the Lorentz symmetry breaking increases the angular width of the glory peak.

\section{Quasinormal modes for the acoustic metric}
\label{S4}
The quasinormal modes are solutions of perturbation equations that satisfy specific conditions. The solution \eqref{condAssintotic1} is rewritten considering only purely ingoing waves at the horizon and purely outgoing waves at spatial infinity \cite{berti2009quasinormal}, in such form
\begin{eqnarray}
\mathcal{R}_{\omega l}  \sim e^{\pm i\tilde{\omega}x}, \qquad (x \rightarrow  \pm \infty).
\label{condQNM}
\end{eqnarray}
There is a discrete spectrum of quasinormal frequencies $\omega_{n}$ where $n$ is the overtone number. These frequencies are complex, where the real part determines the oscillation frequency and the imaginary part determines the damping rate.
\subsection{WKB approximation}
In this work, we will use the WKB approximation to obtain the quasinormal frequencies. This method is widely used in the literature because it presents very satisfactory results. {However, for the canonical acoustic black hole case, the term of ($1/r^{4}$) in the function $\Lambda(r)$ in \eqref{eqRad1} makes the precise calculation of the quasinormal frequencies difficult, not only for the WKB method but also for other methods as discussed in \citep{dolan2010quasinormal}.} Thus, we shall now be using the sixth-order corrected approximation introduced by Konoplya \citep{konoplya2003quasinormal}, which can be written in the form
\begin{eqnarray}
\dfrac{i\left(\omega_{n}^{2} - V_{0}\right)}{\sqrt{-2V''_{0}}} -\sum_{j=2}^{6} \Omega_{j} = n + \dfrac{1}{2},
\end{eqnarray}
where $V_{0}$ is the effective potential at the maximum point. 
Here the quotation marks ($''$) correspond to the second derivative concerning the tortoise coordinate and $\Omega_{j}$ are the corrections.
The results obtained for the quasinormal frequencies $\omega_{n}$ can be verified in Table \eqref{tab1} for some values of the parameter $\alpha$. The quasinormal modes for a usual canonical black hole ($\alpha = 0$) can be checked in \citep{berti2004quasinormal, dolan2010quasinormal}. 

\begin{figure}[!htb]
    \centering
    \subfigure[]{\includegraphics[scale=0.35]{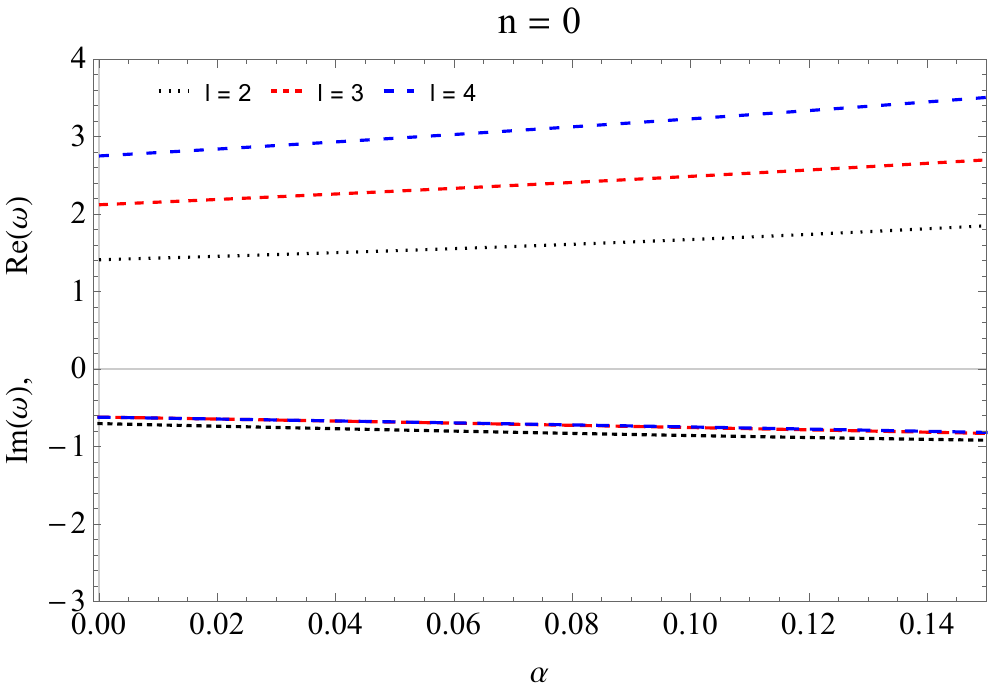}\label{wvsalphan0}}
    \qquad
    \subfigure[]{\includegraphics[scale=0.35]{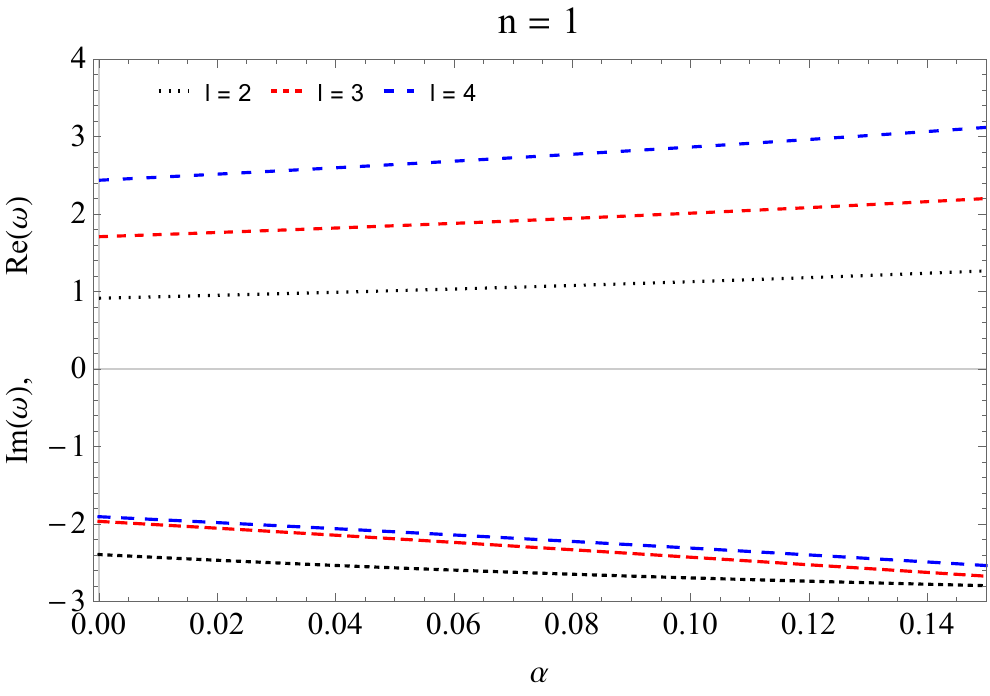}\label{wvsalphan1}}
    \caption{\footnotesize{Quasinormal frequency behavior as a function of the Lorentz symmetry breaking term $\alpha$.}}
    \label{wvsalpha}
\end{figure}

In Table \eqref{tab1}, we see the results for values of $l = 1, 2, 3$, and $4$ for overtone numbers from zero to four. See that the effect of the Lorentz symmetry breaking represented by the parameter $\alpha$ causes a small increase in the imaginary part, such that we have an increase in frequency and damping rate. 
This behavior can be seen in Figure \ref{wvsalpha} for fundamental mode $n = 0$ and first harmonic $n=1$. Note that the change in the imaginary part of the frequency concerning the $\alpha$ parameter is very subtle, especially for the fundamental mode. This effect of reducing the damping rate due to Lorentz symmetry breaking can also be verified in the time domain analysis --- See Figure \ref{QNM_MDF}.

\begin{figure}[!htb]
    \centering
    \includegraphics[scale=0.4]{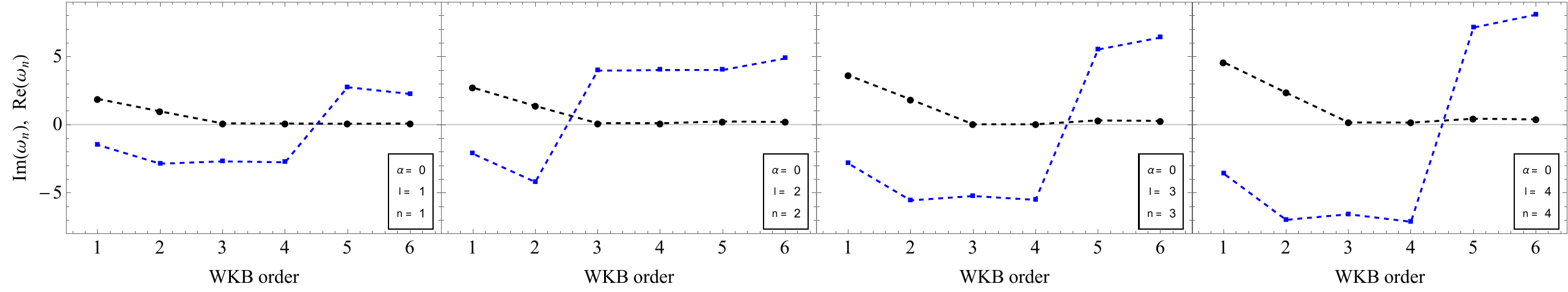}
    \caption{\footnotesize{Real (black) and imaginary (blue) parts of the quasinormal frequency as a function of the WKB order for $l=n$ with $\alpha=0$.}}
    \label{orderWKB} 
\end{figure}

 \begin{table}[ht!]
	\begin{footnotesize}
	\begin{center}
	\caption{\footnotesize{Quasinormal frequencies 6th order WKB.}} 
	\label{tab1}	
	\begin{tabular}{|c||c||c|c|c|c|c}
	\hline
 $ \alpha $ & $ n $   &  $l = 1$ &   $l = 2$ &  $l = 3$ &  $l = 4$  \\
	\hline
	\multirow{5}{*}{$ 0.00 $}
   & 0  & 1.09712 - 0.39365i & 1.41382 - 0.70152i & 2.12332 - 0.61734i & 2.75319 - 0.61870i   \\
   & 1  & 0.09467 + 2.26103i & 0.91744 - 2.39155i & 1.71099 - 1.96392i & 2.43969 - 1.90336i   \\
   & 2  &                    & 0.20105 + 4.86455i & 0.86997 - 3.76131i & 1.77142 - 3.39678i   \\
   & 3  &                    &                    & 0.27826 + 6.40021i & 0.78650 - 5.36918i   \\
   & 4  &                    &                    &                    & 0.39336 + 8.07216i   \\
	\hline
	\hline
	\multirow{5}{*}{$ 0.05 $}
   & 0  & 1.22294 - 0.38039i & 1.53054 - 0.78306i & 2.29856 - 0.68210i & 2.98212 - 0.67954i   \\
   & 1  & 0.01797 - 2.54622i & 1.01435 - 2.56340i & 1.85286 - 2.18854i & 2.64291 - 2.09833i   \\
   & 2  &                    & 0.19046 + 5.31523i & 0.97935 - 4.20619i & 1.93339 - 3.76744i   \\
   & 3  &                    &                    & 0.20139 + 7.17768i & 0.91111 - 5.98655i   \\
   & 4  &                    &                    &                    & 0.29068 + 9.04291i   \\
	\hline
	\hline
	\multirow{5}{*}{$ 0.10 $}
   & 0  & 1.32247 - 0.39275i & 1.67457 - 0.85563i & 2.49010 - 0.75241i & 3.23220 - 0.74499i   \\
   & 1  & 0.15458 - 2.86036i & 1.13029 - 2.69339i & 2.01475 - 2.42633i & 2.86831 - 2.30835i   \\
   & 2  &                    & 0.13682 + 5.75824i & 1.11562 - 4.65864i & 2.12448 - 4.16317i   \\
   & 3  &                    &                    & 0.08809 + 7.95771i & 1.07757 - 6.63336i   \\
   & 4  &                    &                    &                    & 0.13003 + 10.0426i   \\
	\hline
	\hline
	\multirow{5}{*}{$ 0.15 $}
   & 0  & 1.39526 - 0.43675i & 1.85191 - 0.91583i & 2.70316 - 0.82799i & 3.50862 - 0.81548i   \\
   & 1  & 0.31297 - 3.20394i & 1.26994 - 2.79396i & 2.20395 - 2.67153i & 3.12223 - 2.53391i   \\
   & 2  &                    & 0.01687 + 6.19285i & 1.28610 - 5.10483i & 2.35309 - 4.58134i   \\
   & 3  &                    &                    & 0.07432 - 8.71502i & 1.29669 - 7.29887i   \\
   & 4  &                    &                    &                    & 0.10291 - 11.0447i   \\
	\hline
 	\end{tabular}
	\end{center}
	\end{footnotesize}
\end{table}
The positive sign in the imaginary part could suggest an instability in the black hole, but as we mentioned earlier the term of $1/r^{4}$ makes the results inaccurate. In Figure \ref{orderWKB}, we can see that there is an inversion of the sign of the imaginary part (curves in blue) for corrections at large orders, which is possibly caused by the derivatives of the potential. This problem is smoothed out when we increase the value of the parameter $\alpha$, as we can see in the table ($\alpha=0.1 \quad \text{and} \quad 0.15$)

\begin{figure}[!htb]
    \centering
    \includegraphics[scale=0.35]{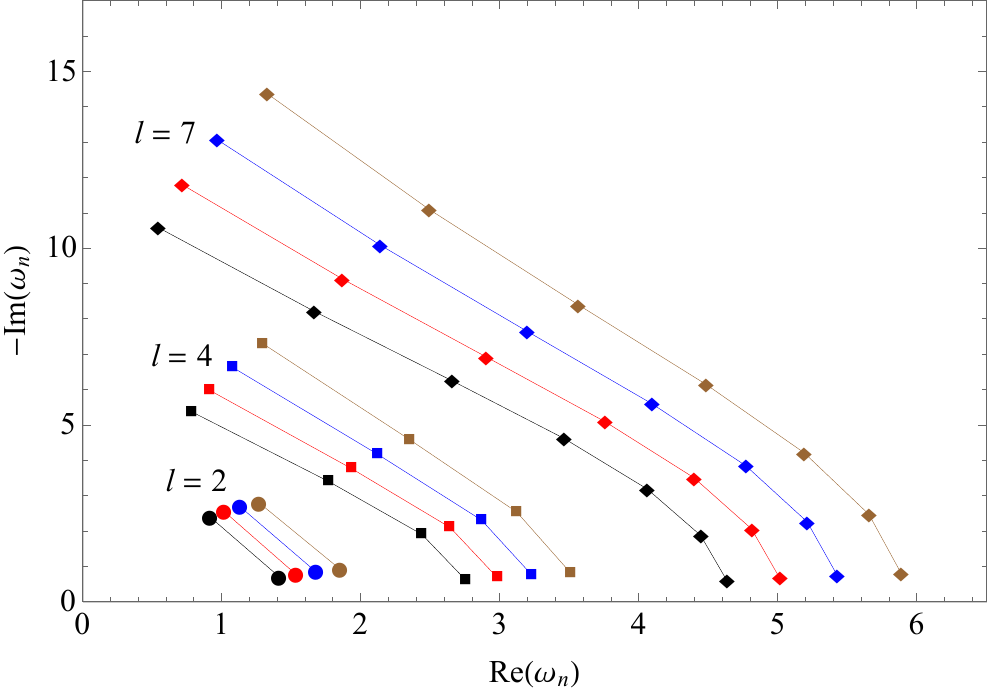}
    \caption{\footnotesize{Complex plan for quasinormal frequencies, considering $\alpha = 0$ (black), $\alpha = 0.05$(red), $\alpha = 0.10$ (blue) and $\alpha = 0.15$ (brown).}}
    \label{ComplexPlan}
\end{figure}

In Figure \ref{ComplexPlan}, we have the results for the quasinormal frequencies in the form of a complex plane, considering three families of multipoles $l = 2, 4$, and $7$. For the first values of the harmonics, the imaginary part grows with the increase of $n$ while the real part decreases. The results in black represent the canonical usual case $(\alpha = 0)$, the effect of the parameter $\alpha$ causes a shift of the curves to the left.

	\subsection{Time-Domain evolution}
\begin{figure}[!htb]
    \centering
    \subfigure[]{\includegraphics[scale=0.32]{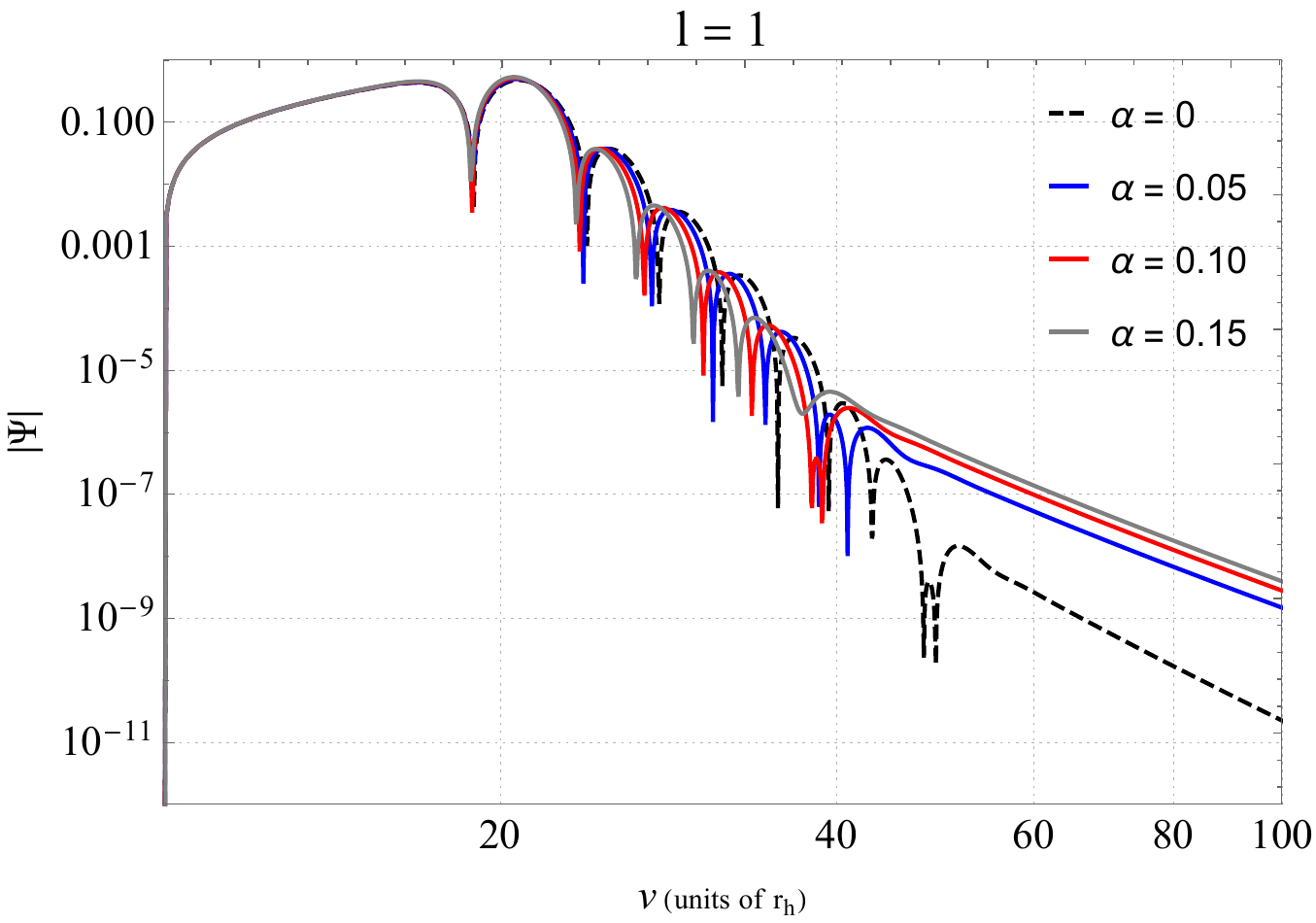}\label{QNM_MDF_(3+1)l1}}
    \qquad
    \subfigure[]{\includegraphics[scale=0.32]{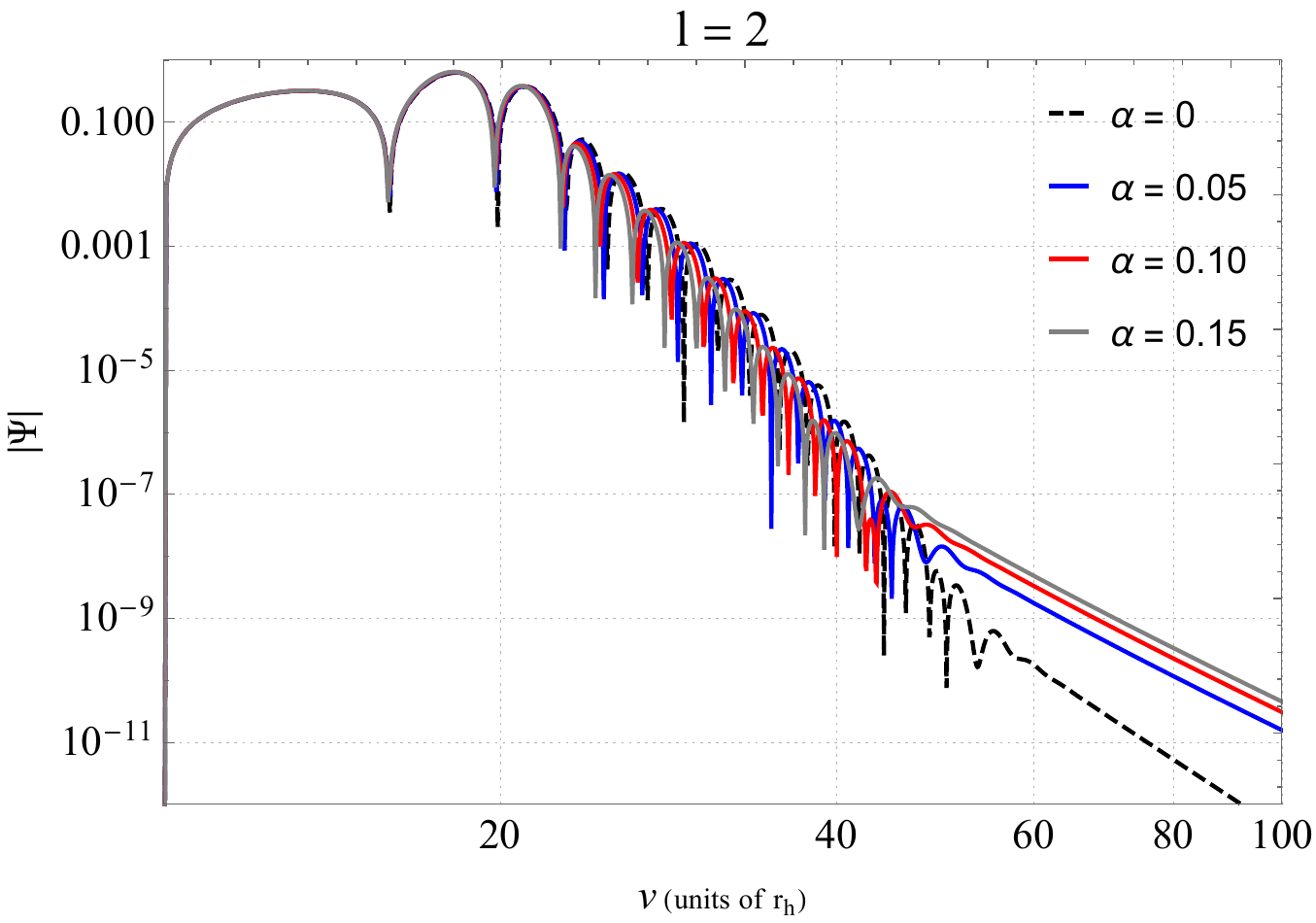}\label{QNM_MDF_(3+1)l2}}
    \caption{\footnotesize{Log-log plots for the wave function at $x=10r_{h}$, comparing the usual case for the canonical acoustic black hole $\alpha = 0$ (dotted line) with the effects.}}
    \label{QNM_MDF}
\end{figure}

We will complement the study, verifying the behavior of the quasinormal modes evolving with time, thus, investigating the scalar disturbances in the temporal domain. The obtained results from the temporal domain represent the frequencies of the fundamental quasinormal modes $n = 0$, that is, less damped.
We begin by writing the wave equation as follows
\begin{eqnarray}
\dfrac{\partial^{2}\mathcal{R}}{\partial t^{2}} - \dfrac{\partial^{2}\mathcal{R}}{\partial x^{2}} + V_{eff}\mathcal{R} = 0.
\label{eRtime}
\end{eqnarray}
The integration technique of the wave equation \eqref{eRtime} was initially developed by Gundlach \cite{gundlach1994late}. Introducing the null coordinates $u = t - x$ and $v = t + x$, the equation can be reduced to the form
\begin{eqnarray}
-4\dfrac{\partial^{2}\mathcal{R}(u,v)}{\partial u\partial v} = V(u,v)\mathcal{R}(u,v).
\label{eqRcone}
\end{eqnarray}
We simulate the evolution of the perturbations by using the two-dimensional finite difference method. The equation \eqref{eqRcone} can be integrated in the form
\begin{eqnarray}
\mathcal{R}(u+h,v+h) = -\mathcal{R}(u,v)+\mathcal{R}(u+h,v)+\mathcal{R}(u,v+h)-\dfrac{h^{2}}{8}V(u,v)\left[\mathcal{R}(u+h,v)+\mathcal{R}(u,v+h)\right] + \mathcal{O}(h^{2}),
\end{eqnarray}
where $h$ is the step length for $u$ and $v$. The equation allows us to calculate the values of the wave function $\mathcal{R}(u,v)$ within a region, built on two null surfaces $u = u_{0}$ and $v = v_{0}$.
We then impose a Gaussian initial condition centered at $v = \bar{v}$ and width $\sigma$ at $u = u_{0}$ of the form
 $\mathcal{R}(u=u_{0},v)= A e^{-(v-\bar{v})/2\sigma^{2}}$ and a second condition $\mathcal{R}(u,v=v_{0}) = 0$. To obtain the results shown in the Figure \ref{QNM_MDF}, we used the following values for the Gaussian parameters, a width $\sigma = 1$ centered on $\bar{v} = 10r_{h}$ and an amplitude $A=1$. We defined an interval of values for $u$ and $v$ between $0$ and $100r_{h}$ with the step in this interval of $h=0.05r_{h}$. This adjustment was enough to obtain a good result as shown below.
In Figure \ref{QNM_MDF} we have the results obtained for the time evolution over a canonical acoustic black hole with two families of multipoles $l=1, 2$. We can see that the behavior of the quasinormal frequencies matches the results obtained by the WKB method shown in Table \ref{tab1}. As already mentioned, the real part is given by the oscillation frequency and the imaginary part by the decay exponent

\subsection{Acoustic shadow radius}
\label{S5}
\begin{figure}[!htb]
    \centering
    \subfigure[]{\includegraphics[scale=0.41]{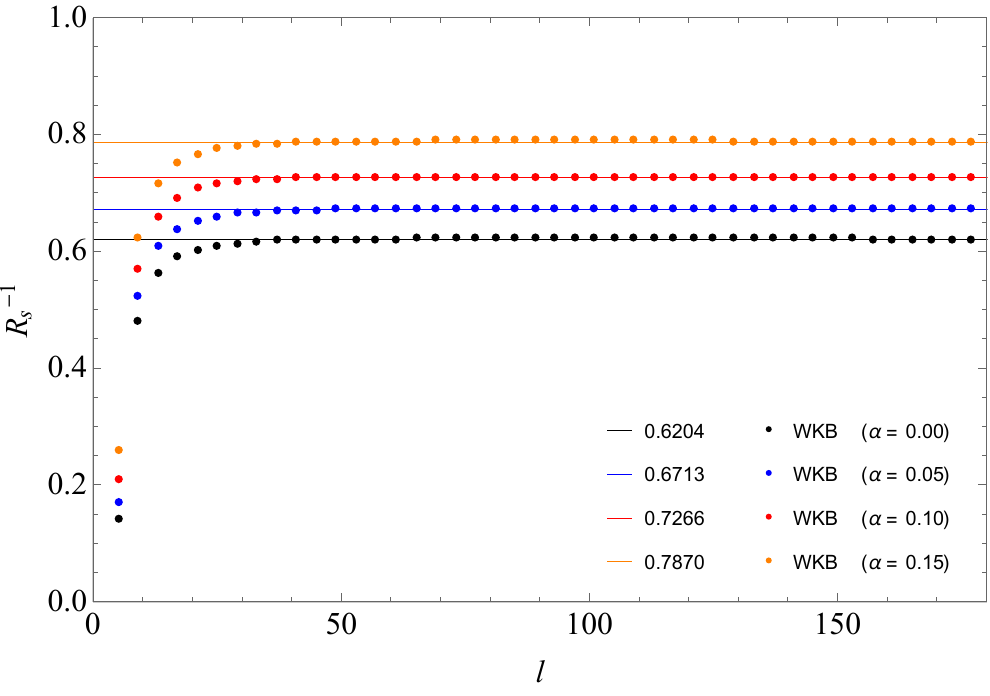}\label{WKBeShadow}}
    \qquad
    \subfigure[]{\includegraphics[scale=0.33]{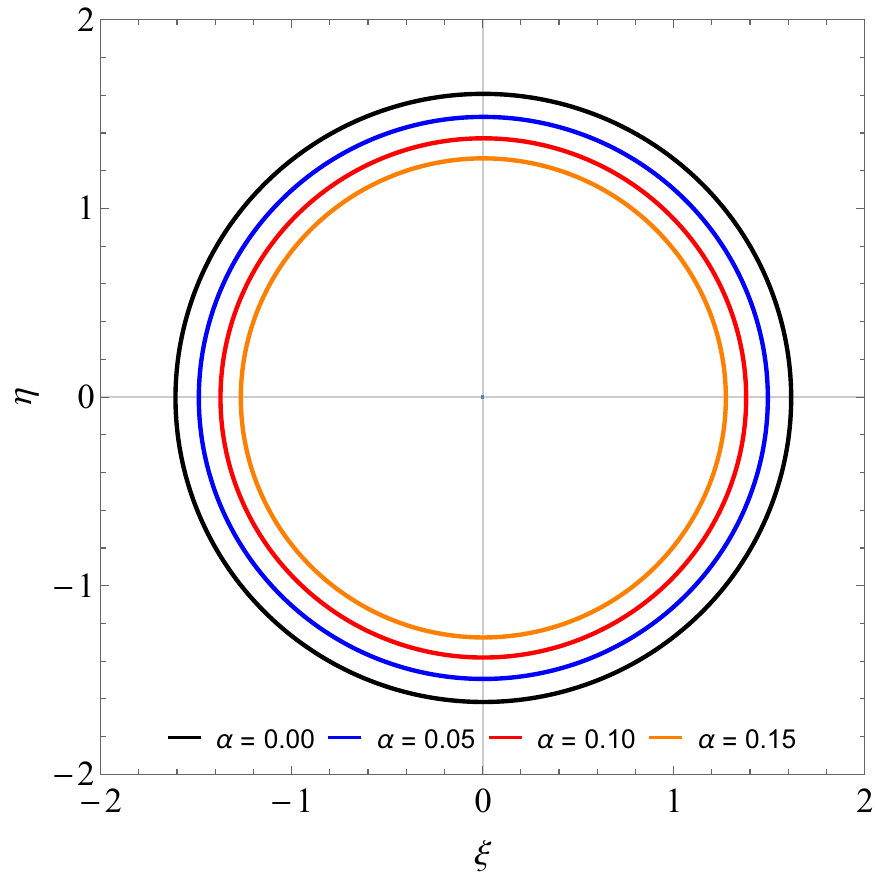}\label{raioShadow}}
    \caption{\footnotesize{In (a) the lines represent the values of each shadow radius for the respective values of $\alpha$. The results of the real part of the quasinormal frequency obtained by the WKB approximation converge to the value of the inverse of the radius for large values of $l$. In (b) we have the representation of the shadow rays for each value of $\alpha$.}}
    \label{Shadow}
\end{figure}

The shadow of the black hole is related to the absorption of photons near the horizon. This spot in the line of sight of an observer, is one of the fingerprints of the geometry around the horizon of the black hole. 
In the acoustic scenario, the shadow of the black hole describes the properties of sound waves near the acoustic horizon, studied in \citep{guo2020acoustic, ling2021shadow}. In the present study, we will verify the effects of Lorentz symmetry breaking for the acoustic black hole shadow.
The shadow for the black hole can be obtained by introducing the so-called celestial coordinates \cite{vazquez2003strong} given by
\begin{eqnarray}
\xi = \lim_{r_{o} \rightarrow \infty}\left[-r_{o}^{2}\sin\theta_{o}\dfrac{d\phi}{dr}\Big |_{\theta = \theta_{o}} \right], \qquad 
\eta = \lim_{r_{o} \rightarrow \infty}\left[r_{o}^{2}\dfrac{d\theta}{dr}\Big |_{\theta = \theta_{o}} \right],
\end{eqnarray}
where $(r_{o},\theta_{o})$ is the position of the observer at infinity. For an observer on an equatorial plane, this is $\theta_{o} = \pi/2$, in our case we have the following value for the shadow radius
\begin{eqnarray}
\label{as}
R_{s} = \sqrt{\xi^{2} + \eta^{2}} = b_{c} = \dfrac{\sqrt{1-2\alpha r_{h}^{2}u_{c}^{2}}}{\sqrt{(1+\alpha)\left(1 - r_{h}^{4}u_{c}^{4}\right)}u_{c}}.
\end{eqnarray}
For $\alpha=0$ the above result is reduced to the shadow radius of the usual canonical acoustic black hole, i.e.
\begin{eqnarray}
R_{s}=\left[\frac{3\sqrt{3}}{2}\right]^{1/2}r_h.
\end{eqnarray}
The relationship between the quasinormal modes and the shadow radius of the canonical black hole can be found in the limit $l>>n$, as initially done by Cardoso \cite{cardoso2009geodesic}, relating the real part of the quasinormal frequency to the shadow radius.
In Fig.~\ref{WKBeShadow}, we have the real part of the quasinormal frequency represented by the points on the graph for several values of multipoles in an interval of $1<l<200$ and the obtained values converge to the inverse of the shadow radius. The behavior of the shadow radius for each value of $\alpha$ can also be observed in the ($\eta,\xi$) plane, see Fig.~\ref{raioShadow}. 

\section{Conclusions.}
\label{S6}
In this paper we study the scattering, absorption and quasinormal modes for a (3+1) dimensions acoustic metric, which describes the canonical black hole for the Abelian Higgs model with the Lorentz-violating term.
We verified that the violation term modifies the absorption cross section both at low and high frequencies, producing a reduction of the absorption cross section. Solving numerically we obtain the complete frequency regime where we confirm the results in the previously mentioned limits. The scattering cross section was obtained using the semi-classical method by calculating the deflection angle. We found that at small angles the parameter $\alpha$ has a small influence. At higher angles we verify numerically that the interference fringes for the scattering cross section shift due to the violation term.
The quasinormal modes were studied through the use of the sixth order WKB approximation and we verified an increase in the real and imaginary frequencies with an increase of $l$ for each value of $\alpha$.
We also conclude that the Lorentz-violating parameter reduces the shadow radius, decreasing the absorption cross section and that in the eikonal limit the relationship between the shadow radius and the real part of the quasinormal frequency is preserved.

{\acknowledgments} 
We would like to thank CNPq, CAPES and CNPq/PRONEX/FAPESQ-PB (Grant nos. 165/2018 and 015/2019),
for partial financial support. MAA, FAB and EP acknowledge support from CNPq (Grant nos. 306398/2021-4,
309092/2022-1, 304290/2020-3). JAVC would also like to thank FAPESQ-PB/CNPq n$^0$ 77/2022 for financial support.

\begin{thebibliography}{90}

\bibitem{scientific2016tests}
L.~Scientific, V.~Collaborations, B.~Abbott, R.~Abbott, T.~Abbott,
  M.~Abernathy, F.~Acernese, K.~Ackley, C.~Adams, T.~Adams, {\em et~al.},
  ``\textit{Tests of general relativity with gw150914},'' {Phys. Rev. Lett.},
  \textbf{116}, no.~22, 221101 (2016).

\bibitem{abbott2017gw170817}
B.~P. Abbott, R.~Abbott, T.~Abbott, F.~Acernese, K.~Ackley, C.~Adams, T.~Adams,
  P.~Addesso, R.~Adhikari, V.~B. Adya, {\em et~al.}, ``\textit{Gw170817: observation of
  gravitational waves from a binary neutron star inspiral},'' { Phys.
  Rev. Lett.},\textbf{ 119}, no.~16, 161101 (2017).

\bibitem{akiyama2019firstResultsIV}
K.~Akiyama, A.~Alberdi, W.~Alef, K.~Asada, R.~Azulay, A.-K. Baczko, D.~Ball,
  M.~Balokovi{\'c}, J.~Barrett, D.~Bintley, {\em et~al.}, ``\textit{First m87 event
  horizon telescope results. iv. imaging the central supermassive black hole},''
  Astrophys. J. Lett. \textbf{875}, no.1, L4 (2019)
doi:10.3847/2041-8213/ab0e85
[arXiv:1906.11241 [astro-ph.GA]].

\bibitem{EventHorizonTelescope:2019ggy}
K.~Akiyama \textit{et al.} [Event Horizon Telescope],
``\textit{First M87 Event Horizon Telescope Results. VI. The Shadow and Mass of the Central Black Hole},''
Astrophys. J. Lett. \textbf{875}, no.1, L6 (2019)
doi:10.3847/2041-8213/ab1141
[arXiv:1906.11243 [astro-ph.GA]].

\bibitem{akiyama2022first}
K.~Akiyama, A.~Alberdi, W.~Alef, J.~C. Algaba, R.~Anantua, K.~Asada, R.~Azulay,
  U.~Bach, A.-K. Baczko, D.~Ball, {\em et~al.}, 
``\textit{First Sagittarius A* Event Horizon Telescope Results. I. The Shadow of the Supermassive Black Hole in the Center of the Milky Way},''
Astrophys. J. Lett. \textbf{930}, no.2, L12 (2022)
doi:10.3847/2041-8213/ac6674

\bibitem{unruh1981experimental}
W.~G. Unruh, ``\textit{Experimental black hole evaporation},''
Phys. Rev. Lett. \textbf{46}, 1351-1353 (1981)
doi:10.1103/PhysRevLett.46.1351

\bibitem{ge2010acoustic}
X.-H. Ge and S.-J. Sin,
``\textit{Acoustic black holes for relativistic fluids},''
JHEP \textbf{06}, 087 (2010)
doi:10.1007/JHEP06(2010)087
[arXiv:1001.0371 [hep-th]].

\bibitem{ge2019acoustic}
X.-H. Ge, M.~Nakahara, S.-J. Sin, Y.~Tian, and S.-F. Wu,
``\textit{Acoustic black holes in curved spacetime and the emergence of analogue Minkowski spacetime},''
Phys. Rev. D \textbf{99}, no.10, 104047 (2019)
doi:10.1103/PhysRevD.99.104047
[arXiv:1902.11126 [hep-th]].

\bibitem{anacleto2010acoustic}
M.~A.~Anacleto, F.~A.~Brito and E.~Passos,
``\textit{Acoustic Black Holes from Abelian Higgs Model with Lorentz Symmetry Breaking},''
Phys. Lett. B \textbf{694}, 149-157 (2011)
doi:10.1016/j.physletb.2010.09.045
[arXiv:1004.5360 [hep-th]].

\bibitem{anacleto2011superresonance}
M.~A.~Anacleto, F.~A.~Brito and E.~Passos,
``\textit{Superresonance effect from a rotating acoustic black hole and Lorentz symmetry breaking},''
Phys. Lett. B \textbf{703}, 609-613 (2011)
doi:10.1016/j.physletb.2011.08.040
[arXiv:1101.2891 [hep-th]].


\bibitem{Anacleto:2023ali}
M.~A.~Anacleto, F.~A.~Brito and E.~Passos,
``\textit{Modified metrics of acoustic black holes: A review},''
Phys. Sci. \& Biophys. J. \textbf{7}, 000245 (2023)
doi:10.23880/psbj-16000245
[arXiv:2306.03077 [hep-th]].

\bibitem{Anacleto:2011bv}
M.~A.~Anacleto, F.~A.~Brito and E.~Passos,
``\textit{Supersonic Velocities in Noncommutative Acoustic Black Holes},''
Phys. Rev. D \textbf{85}, 025013 (2012)
doi:10.1103/PhysRevD.85.025013
[arXiv:1109.6298 [hep-th]].

\bibitem{Anacleto:2012du}
M.~A.~Anacleto, F.~A.~Brito and E.~Passos,
``\textit{Noncommutative analogue Aharonov-Bohm effect and superresonance},''
Phys. Rev. D \textbf{87}, no.12, 125015 (2013)
doi:10.1103/PhysRevD.87.125015
[arXiv:1210.7739 [hep-th]].

\bibitem{Anacleto:2021nhm}
M.~A.~Anacleto, F.~A.~Brito, G.~C.~Luna and E.~Passos,
``\textit{The generalized uncertainty principle effect in acoustic black holes},''
Annals Phys. \textbf{440}, 168837 (2022)
doi:10.1016/j.aop.2022.168837
[arXiv:2112.13573 [gr-qc]].

\bibitem{Anacleto:2013esa}
M.~A.~Anacleto, F.~A.~Brito and E.~Passos,
``\textit{Acoustic Black Holes and Universal Aspects of Area Products},''
Phys. Lett. A \textbf{380}, 1105-1109 (2016)
doi:10.1016/j.physleta.2016.01.030
[arXiv:1309.1486 [hep-th]].

\bibitem{Casalderrey-Solana:2004fdk}
J.~Casalderrey-Solana, E.~V.~Shuryak and D.~Teaney,
``\textit{Conical flow induced by quenched QCD jets},''
J. Phys. Conf. Ser. \textbf{27}, 22-31 (2005)
doi:10.1088/1742-6596/27/1/003
[arXiv:hep-ph/0411315 [hep-ph]].

\bibitem{Das:2020zah}
A.~Das, S.~S.~Dave, O.~Ganguly and A.~M.~Srivastava,
``\textit{Hawking radiation from acoustic black holes in relativistic heavy ion collisions},''
Phys. Lett. B \textbf{817}, 136294 (2021)
doi:10.1016/j.physletb.2021.136294
[arXiv:2006.15912 [gr-qc]].

\bibitem{GarciadeAndrade:2008kz}
L.~C.~Garcia de Andrade,
``\textit{Kerr-Schild Riemannian acoustic black holes in dynamo plasma laboratory},''
[arXiv:0808.2271 [gr-qc]].


\bibitem{Ditta:2023lny}
A.~Ditta, T.~Xia and M.~Yasir,
``Particle motion and lensing with plasma of acoustic Schwarzschild black hole,''
Int. J. Mod. Phys. A \textbf{38}, no.06n07, 2350041 (2023)
doi:10.1142/S0217751X23500410


\bibitem{regge1957stability}
T.~Regge and J.~A.~Wheeler, 
``\textit{Stability of a Schwarzschild singularity},''
Phys. Rev. \textbf{108}, 1063-1069 (1957)
doi:10.1103/PhysRev.108.1063

\bibitem{kokkotas1999quasi}
K.~D. Kokkotas and B.~G. Schmidt, 
``\textit{Quasinormal modes of stars and black holes},''
Living Rev. Rel. \textbf{2}, 2 (1999)
doi:10.12942/lrr-1999-2
[arXiv:gr-qc/9909058 [gr-qc]].

\bibitem{schutz1985black}
B.~F. Schutz and C.~M. Will, 
``\textit{Black hole normal modes: A semianalytic approach},''
Astrophys. J. Lett. \textbf{291}, L33-L36 (1985)
doi:10.1086/184453

\bibitem{iyer1987black}
S.~Iyer and C.~M. Will, 
``\textit{Black Hole Normal Modes: A {WKB} Approach. 1. Foundations and Application of a Higher Order {WKB} Analysis of Potential Barrier Scattering},''
Phys. Rev. D \textbf{35}, 3621 (1987)
doi:10.1103/PhysRevD.35.3621

\bibitem{konoplya2003quasinormal}
R.~Konoplya, 
``\textit{Quasinormal behavior of the d-dimensional Schwarzschild black hole and higher order WKB approach},''
Phys. Rev. D \textbf{68}, 024018 (2003)
doi:10.1103/PhysRevD.68.024018
[arXiv:gr-qc/0303052 [gr-qc]].

\bibitem{leaver1985analytic}
E.~W. Leaver, 
``\textit{An Analytic representation for the quasi normal modes of Kerr black holes},''
Proc. Roy. Soc. Lond. A \textbf{402}, 285-298 (1985)
doi:10.1098/rspa.1985.0119

\bibitem{leaver1990quasinormal}
E.~W. Leaver, ``\textit{Quasinormal modes of Reissner-Nordstrom black holes},''
Phys. Rev. D \textbf{41}, 2986-2997 (1990)
doi:10.1103/PhysRevD.41.2986

\bibitem{visser1998acoustic}
M.~Visser, 
``\textit{Acoustic black holes: Horizons, ergospheres, and Hawking radiation},''
Class. Quant. Grav. \textbf{15}, 1767-1791 (1998)
doi:10.1088/0264-9381/15/6/024
[arXiv:gr-qc/9712010 [gr-qc]].

\bibitem{berti2004quasinormal}
E.~Berti, V.~Cardoso, and J.~P. Lemos, 
``\textit{Quasinormal modes and classical wave propagation in analogue black holes},''
Phys. Rev. D \textbf{70}, 124006 (2004)
doi:10.1103/PhysRevD.70.124006
[arXiv:gr-qc/0408099 [gr-qc]].

\bibitem{cardoso2004quasinormal}
V.~Cardoso, J.~P. Lemos, and S.~Yoshida, 
``\textit{Quasinormal modes and stability of the rotating acoustic black hole: Numerical analysis},''
Phys. Rev. D \textbf{70}, 124032 (2004)
doi:10.1103/PhysRevD.70.124032
[arXiv:gr-qc/0410107 [gr-qc]].

\bibitem{mohr2008codata}
P.~J. Mohr, B.~N. Taylor, and D.~B. Newell, 
``\textit{CODATA Recommended Values of the Fundamental Physical Constants: 2006},''
Rev. Mod. Phys. \textbf{80}, 633-730 (2008)
doi:10.1103/RevModPhys.80.633
[arXiv:0801.0028 [physics.atom-ph]].

\bibitem{Casana:2011bv}
R.~Casana and K.~A.~T.~da Silva,
Mod. Phys. Lett. A \textbf{30}, no.07, 1550037 (2015)
doi:10.1142/S0217732315500376
[arXiv:1106.5534 [hep-th]].

\bibitem{anacleto2019quantum}
M.~A.~Anacleto, F.~A.~Brito, C.~V.~Garcia, G.~C.~Luna and E.~Passos, 
``\textit{Quantum-corrected rotating acoustic black holes in Lorentz-violating background},''
Phys. Rev. D \textbf{100}, no.10, 105005 (2019)
doi:10.1103/PhysRevD.100.105005
[arXiv:1904.04229 [hep-th]].

\bibitem{collins1973elastic}
P.~A.~Collins, R.~Delbourgo and R.~M.~Williams,
``\textit{On the elastic Schwarzschild scattering cross-section},''
J. Phys. A \textbf{6}, 161-169 (1973)
doi:10.1088/0305-4470/6/2/007

\bibitem{dolan2009scattering}
S.~R.~Dolan, E.~S.~Oliveira and L.~C.~B.~Crispino,
``\textit{Scattering of sound waves by a canonical acoustic hole},''
Phys. Rev. D \textbf{79}, 064014 (2009)
doi:10.1103/PhysRevD.79.064014
[arXiv:0904.0010 [gr-qc]].

\bibitem{oliveira2010absorption}
E.~S.~Oliveira, S.~R.~Dolan and L.~C.~B.~Crispino,
``\textit{Absorption of planar waves in a draining bathtub},''
Phys. Rev. D \textbf{81}, 124013 (2010)
doi:10.1103/PhysRevD.81.124013
 
\bibitem{Anacleto:2020efy}
M.~A.~Anacleto, F.~A.~Brito, B.~R.~Carvalho and E.~Passos,
``\textit{Noncommutative correction to the entropy of BTZ black hole with GUP},''
Adv. High Energy Phys. \textbf{2021}, 6633684 (2021)
doi:10.1155/2021/6633684
[arXiv:2010.09703 [hep-th]].

\bibitem{visser2013area}
M.~Visser, ``\textit{Area products for stationary black hole horizons},''
Phys. Rev. D \textbf{88}, no.4, 044014 (2013)
doi:10.1103/PhysRevD.88.044014
[arXiv:1205.6814 [hep-th]].

\bibitem{abramowitz1965handbook}
M.~Abramowitz and I.~A. Stegun, 
``\textit{Handbook of mathematical functions dover publications},'' 
{New York}, vol.~361, 1965.

\bibitem{morse1954methods}
P.~M. Morse, H.~Feshbach and E.~L.~Hill, 
``\textit{Methods of Theoretical Physics},'' 
Am. J. Phys. \textbf{22}, 410–413 (1954)
doi:10.1119/1.1933765

\bibitem{ford1959semiclassical}
K.~W.~Ford and J.~A.~Wheeler,
``\textit{Semiclassical description of scattering},'' 
Annals Phys. \textbf{7} (3), 259-286 (1959) 
doi:10.1016/0003-4916(59)90026-0

\bibitem{Ford:2000uye}
K.~W.~Ford and J.~A.~Wheeler,
``\textit{Semiclassical Description of Scattering},''
Annals Phys. \textbf{281}, no.1-2, 608-635 (2000)
doi:10.1006/aphy.2000.6018

\bibitem{arfken1999mathematical}
G.~B. Arfken and H.~J. Weber, ``\textit{Mathematical methods for physicists},'' 1999.

\bibitem{crispino2007absorption}
L.~C.~B.~Crispino, E.~S.~Oliveira and G.~E.~A.~Matsas,
``\textit{Absorption cross section of canonical acoustic holes},''
Phys. Rev. D \textbf{76}, 107502 (2007)
doi:10.1103/PhysRevD.76.107502

\bibitem{Anacleto:2017kmg}
M.~A.~Anacleto, F.~A.~Brito, S.~J.~S.~Ferreira and E.~Passos,
``\textit{Absorption and scattering of a black hole with a global monopole in f(R) gravity},''
Phys. Lett. B \textbf{788}, 231-237 (2019)
doi:10.1016/j.physletb.2018.11.020
[arXiv:1701.08147 [hep-th]].
  
\bibitem{anacleto2020absorption}
M.~A.~Anacleto, F.~A.~Brito, J.~A.~V.~Campos and E.~Passos,
``\textit{Absorption and scattering of a noncommutative black hole},''
Phys. Lett. B \textbf{803}, 135334 (2020)
doi:10.1016/j.physletb.2020.135334
[arXiv:1907.13107 [hep-th]].

\bibitem{Anacleto:2020zhp}
M.~A.~Anacleto, F.~A.~Brito, J.~A.~V.~Campos and E.~Passos,
``\textit{Absorption and scattering by a self-dual black hole},''
Gen. Rel. Grav. \textbf{52}, no.10, 100 (2020)
doi:10.1007/s10714-020-02756-1
[arXiv:2002.12090 [hep-th]].

\bibitem{Anacleto:2020lel}
M.~A.~Anacleto, F.~A.~Brito, J.~A.~V.~Campos and E.~Passos,
``\textit{Quantum-corrected scattering and absorption of a Schwarzschild black hole with GUP},''
Phys. Lett. B \textbf{810}, 135830 (2020)
doi:10.1016/j.physletb.2020.135830
[arXiv:2003.13464 [gr-qc]].

\bibitem{Anacleto:2022shk}
M.~A.~Anacleto, F.~A.~Brito, J.~A.~V.~Campos and E.~Passos,
``\textit{Absorption, scattering and shadow by a noncommutative black hole with global monopole},''
Eur. Phys. J. C \textbf{83}, no.4, 298 (2023)
doi:10.1140/epjc/s10052-023-11484-0
[arXiv:2212.13973 [hep-th]].

\bibitem{matzner1985glory}
R.~A.~Matzner, C.~DeWitte-Morette, B.~Nelson and T.~R.~Zhang,
``\textit{Glory scattering by black holes},''
Phys. Rev. D \textbf{31}, no.8, 1869 (1985)
doi:10.1103/PhysRevD.31.1869

\bibitem{slater1965handbook}
L.~Slater, M.~Abramowitz, and I.~Stegun, ``\textit{Handbook of mathematical
  functions},'' {Abramowitz and IA Stegun, Eds.(US Govt. Printing Office,
  Washington, DC, 1968) Appl. Math. Ser}, vol.~55, 1965.
  
\bibitem{berti2009quasinormal}
E.~Berti, V.~Cardoso and A.~O.~Starinets,
``\textit{Quasinormal modes of black holes and black branes},''
Class. Quant. Grav. \textbf{26}, 163001 (2009)
doi:10.1088/0264-9381/26/16/163001
[arXiv:0905.2975 [gr-qc]].

\bibitem{dolan2010quasinormal}
S.~R.~Dolan, L.~A.~Oliveira and L.~C.~B.~Crispino,
``\textit{Quasinormal modes and Regge poles of the canonical acoustic hole},''
Phys. Rev. D \textbf{82}, 084037 (2010)
doi:10.1103/PhysRevD.82.084037
[arXiv:1407.3904 [gr-qc]].

\bibitem{gundlach1994late}
C.~Gundlach, R.~H.~Price and J.~Pullin,
``\textit{Late time behavior of stellar collapse and explosions: 1. Linearized perturbations},''
Phys. Rev. D \textbf{49}, 883-889 (1994)
doi:10.1103/PhysRevD.49.883
[arXiv:gr-qc/9307009 [gr-qc]].
 
\bibitem{guo2020acoustic}
H.~Guo, H.~Liu, X.~M.~Kuang and B.~Wang,
``\textit{Acoustic black hole in Schwarzschild spacetime: quasi-normal modes, analogous Hawking radiation and shadows},''
Phys. Rev. D \textbf{102}, 124019 (2020)
doi:10.1103/PhysRevD.102.124019
[arXiv:2007.04197 [gr-qc]].

\bibitem{ling2021shadow}
R.~Ling, H.~Guo, H.~Liu, X.~M.~Kuang and B.~Wang,
``\textit{Shadow and near-horizon characteristics of the acoustic charged black hole in curved spacetime},''
Phys. Rev. D \textbf{104}, no.10, 104003 (2021)
doi:10.1103/PhysRevD.104.104003
[arXiv:2107.05171 [gr-qc]].

\bibitem{vazquez2003strong}
S.~E.~Vazquez and E.~P.~Esteban,
``\textit{Strong field gravitational lensing by a Kerr black hole},''
Nuovo Cim. B \textbf{119}, 489-519 (2004)
doi:10.1393/ncb/i2004-10121-y
[arXiv:gr-qc/0308023 [gr-qc]].

\bibitem{cardoso2009geodesic}
V.~Cardoso, A.~S.~Miranda, E.~Berti, H.~Witek and V.~T.~Zanchin,
``\textit{Geodesic stability, Lyapunov exponents and quasinormal modes},''
Phys. Rev. D \textbf{79}, no.6, 064016 (2009)
doi:10.1103/PhysRevD.79.064016
[arXiv:0812.1806 [hep-th]].


\end{thebibliography}

\end{document}